\documentstyle[11pt,epsf]{article}
cm \textheight 22.0 true cm \headheight 0 cm \headsep 0 cm
\begin{document}

\begin{center}

{\Large \textbf{Quantum Field Theory and Differential Geometry }}

\vspace{5mm}
 W.F. Chen

 \vspace{2mm}

{\small Department of Physics, University of Winnipeg, Winnipeg,
Manitoba, Canada R3B 2E9}

\end{center}

\begin{abstract}
\noindent We introduce the historical development and physical
idea behind topological Yang-Mills theory and explain how a
physical framework describing subatomic physics  can be used as a
tool to study differential geometry. Further, we emphasize that
this phenomenon demonstrates that the interrelation between
physics and mathematics have come into a new stage.
\end{abstract}

\section{Introduction}
\label{sect1}

In the past 20th century  the interrelation between mathematics and
theoretical high energy physics had gone through an unprecedented
revolution. The traditional viewpoint that mathematics is a language
tool of describing a physical law and a physical problem stimulates
the development in mathematics has been changed considerably.
Mathematics, especially one of its branches, differential geometry,
has begun to merge together with the theoretical framework
describing subatomic physics. The effect of theoretical physics on
the advancement in mathematics is no longer indirect and unilateral.
It is well known that  Einstein's theory of general relativity had
ever greatly promoted the study on Riemannian geometry. But now it
is not the case any more, and it is rather than a physical principle
is becoming a tool of studying mathematics. A remarkable phenomenon
is during the past twenty years  a number of Fields medalists'
works are relevant to theoretical physics. Especially, in 1990 a
leading theoretical physicist, Professor Edward Witten at  the
Institute of Advanced Study in Princeton, was awarded the Fields
Medal  for his pioneering work using a relativistic quantum field
theory to study differential topology of low-dimensional manifold.
This event made a great sensation among both physicists and
mathematicians at that time. These facts imply that the relation
between mathematics and theoretical high energy physics has come to
a new stage and fundamental principles in physics are becoming
one of the necessities in pushing mathematics forward.\\

Naturally people may be puzzled with this phenomenon: mathematics
and physics are two distinct  subjects. Mathematics lays stress on
 rigor and logic, and the development in each step needs a
rigorous proof to support; While  theoretical physics is based on a
scientific hypothesis and  the subsequent experimental test. How can
these two distinct disciplines merge together. The aim of this
article is using topological quantum field theory to introduce and
explain why and how this phenomenon has happened.

\section{Symmetry in Quantum Field Theory and Group Representation}
\label{sect2}

The  interrelation between mathematics and theoretical high energy
physics is an inevitable outcome of modern physics. It is well known
that modern physics started at the beginning of 20th century and
originated from the discovery on  the theory of special relativity
and quantum mechanics. The theory of special relativity is a
fundamental physical principle for the matter at extremely high energy, and
it puts space and time on the same footing. This description has
greatly modified  the traditional space-time version described by
Newtonian mechanics. In fact, the space-time theory described by
Newton's mechanics is only a low-energy approximation to the version
depicted  by the theory of special relativity. Correspondingly, in
mathematics, a mathematician Hermann Minkowski defined a highly
symmetric pseudo-Euclidean space, which is now called the Minkowski
space. Then the theory of special relativity is just realized as an
isometry symmetry in this space, i.e., the distance of the Minkowski
space is invariant under the Lorentz transformation. It is an
elegant geometric description to the theory of special relativity.
On the other hand, the quantum theory is another fundamental
principle for the matter at microscopic scale (atomic and subatomic
size). The atom presents wave behavior and obeys quantum principle
different from that described by classical physics. The classical
physical principle is just the macroscopic limit of quantum theory.
The geometric description on quantum theory is the Hilbert space,
named after the great German mathematician David Hilbert. A physical
state and a physical observable are a vector and a self-adjoint
operator in the Hilbert space, respectively. Further, the
combination of the theory of special relativity and quantum
mechanics had led to the birth of a most powerful theoretical
framework describing the interaction among elementary particles
--- quantum
field theory.\\

The characteristics of quantum field theory  determines that it not
only serves as a theoretical framework describing particle
interactions, but also as  a tool for realizing group representation
and exploring the topology of differential manifold. A quantum field
theory can be constructed from a classical field theory through a
so-called quantization procedure. The basic elements constituting a
classical field theory  are field functions. However, they are not
simply  functions of space-time coordinate, and they have to
constitute certain representations of the Lorentz group (or its
covering group). Thus field functions are classified into scalar,
spinor, vector and higher order tensor according to irreducible
representations of the Lorentz group. In addition, the field
functions may carry other indices other than the space-time ones, so
they can form representations of certain other Lie groups. Further,
the eminent mathematician Hermann Weyl proposed the celebrated gauge
principle for the Abelian $U(1)$ group, and it was later generalized
to the non-Abelian $SU(2)$ group by theoretical physicists Chen-Ning
Yang and Robert Mills \cite{yangm}. The gauge principle states that
if we localize the group representation realized on a field
function, i.e., the representation matrix of the Lie group realized
on a field function being a function of space-time coordinate, then
the group representation index will become a dynamical degree of
freedom and a certain new vector field must be introduced to
preserve the symmetry represented  by the Lie group. In physics, the
newly introduced vector field plays the role of mediating the
interactions between elementary particles. Therefore, gauge
principle is a fundamental principle of constructing a theory
describing the interaction among
elementary particles.\\

The application of quantum field theory in group representation
theory is attributed to the female mathematician Emmy N\"{o}ether.
She discovered  a direct relation between the symmetry of a field
theory and its conservative quantity, and  proposed the celebrated
N\"{o}ether theorem. This theorem shows that the conservative
quantity in a field theory is just a representation for the
generator of the continuous symmetry group. A classical field theory
is described by a functional composed of field functions and their
space-time derivatives not more than second order, which is called a
classical action. The symmetry of a field theory is the invariance
of the classical action under the transformation of the field
function.  In principle, a quantum field theory can be obtained
through a standard procedure
--- either canonical quantization or path integral quantization.
Similar to its classical counterpart, the quantum field theory is
also described by a functional composed of quantum field variables
and their derivatives with respect to  space-time coordinate. This
functional is usually called a quantum effective action, which
consists of the classical action plus quantum corrections. The
quantum effective action is actually the sum of inner products
between two vectors in the Hilbert space of quantum field theory,
and these products are termed as the Green functions, which take
into account the casualty of physical processes. However, due to the
highly non-linear couplings among field functions in the classical
action,  we usually have no way to obtain the precise form of the
quantum effective action. In the case that the coupling is weak, we
can use the perturbative iteration method to calculate the quantum
correction to a certain order of coupling constant. Moreover, one
has to use the regularization and renormalization techniques
invented by theoretical physicists to define the quantum effective
action. In some cases, the quantum effective action may fail to
present certain symmetries that the classical action possesses, this
means certain symmetries become anomalous. If the quantum effective
action still has various symmetries presented in the classical
theory, i.e., no anomaly arises, the Green functions of a quantum
field theory must satisfy certain relations dominated by symmetries,
and these relations are called the Ward identities in gauge
theories. Furthermore, if the vacuum state or ground state (the
state with the lowest energy) of the theory provides a trivial
representation to the symmetry group, which is called no spontaneous
symmetry breaking in physics, then the Hilbert space is a natural
platform providing various representations for the symmetry group.
The quantum states, i.e., vectors in the Hilbert space, are
classified into various irreducible representations of the symmetry
group. In physics the vectors sharing the same irreducible
representation is called constituting a multiplet. The Ward
identities among various Green functions are just reflections of the
symmetry in the field theory system. On the other hand, these
identities have imposed very strong constraints on the quantum
effective action and provide selection rules to the occurrences of
physical processes. Overall, quantum field theory is a natural
physical framework to study group representation due to its own
specifical features, and the physical phenomena it describes is
actually a reflection of symmetry
in nature.\\

\section{Fibre Bundle, Gauge Theory, Instanton Moduli Space
and Instanton Tunneling Effect} \label{sect3}

The connection between gauge theory and the geometry of fibre bundle
is very dramatic. The non-Abelian gauge theory was proposed by
theoretical physicists Chen-Ning Yang and Robert Mills in the early
middle of 1950s \cite{yangm}. At that time, the fibre bundle theory
had already developed ripely in differential geometry, but
physicists almost knew nothing  about it. It was  until early 1960s
that a theoretical physicist Elihu Lubkin  realized that the
classical Yang-Mills gauge theory and the affine geometry of fibre
bundle are identical \cite{lubk}. A gauge field is actually the
pull-back of the connection to the base manifold  of a certain
principle fibre bundle and the gauge field strength is the pull-back
of the curvature of principle bundle, while the gauge group is the
structure group of principle bundle. Further, matter fields can be
considered as the sections of some associated bundles of the
principle fibre bundle, and the gauge transformation is the action
of structure group on the section of associated bundle. However,
these facts had not been taken seriously by physicists. The
extensive application of fibre bundle geometry in gauge theory was
caused by  an article on the global structure of electromagnetic
field written by Tai-Tsun Wu  and Chen-Ning Yang in 1975
\cite{wuyang}. In this article, they defined the electromagnetic
potential on two half spheres $S^2$ with overlapping and  avoided
the singular string problem in the magnetic potential produced by
the Dirac monopole \cite{dirac}. In the overlapping region of two
spheres, the magnetic potentials are related by a $U(1)$  gauge
transformation. The geometry of this physical system is precisely a
 principle $U(1)$-bundle with base manifold $S^2$ \cite{wuyang}. In
the the same year as Wu and Yang published their paper, A.A. Belavin,
A.M. Polyakov, A.S. Schwartz and Yu.S. Tyupkin from the former
Soviet Union found a solution to the classical $SU(2)$ Yang-Mills
theory with finite action in the Euclidean space \cite{bpst}, which
is now called the BPST instanton (due to the physical effect it
produces which will be mentioned later). This solution describes a
gauge field configuration with (anti-)self-dual field strength, and
its geometrical description is the  principle $SU(2)$-bundle on the
base manifold $S^4$. Further, it had been realized that  this
solution has topological meaning, and it is characterized by a
topological index, which is precisely the Chern number of the second
Chern Class in fibre bundle theory \cite{chern1}. The BPST instanton
is actually a classical solution to the Euclidean $SU(2)$ Yang-Mills
theory with the Chern number equal to one. In the following years,
theoretical physicists including Edward Witten \cite{witten6}, E.
Corrigan and D.B. Fairlie \cite{cofa1}, Roman Jackiw, C. Nohl and
Claudio Rebbi \cite{jackiw4} found the multiple (anti-)instanton
solution with Chern number $ |k|
>1$. Especially, Jackiw and Rebbi showed that the classical
Yang-Mills theory  has not only non-Abelian gauge symmetry and the
Poincar\'{e} space-time symmetry, but also a larger conformal
space-time symmetry \cite{jackiw2}, which consists of the
Poincar\'{e} symmetry composed of translational and Lorentz
rotational invariance, dilatational invariance and special conformal
symmetry. As we know, if a field theory has a certain symmetry, then
the symmetry transformation on a space-time dependent classical
solution  should lead to another solution  to the classical equation of
motion. This means that there exists a family of solutions to a
field theory with symmetries. The inequivalent solutions under
symmetry transformations constitute a finite-dimensional space,
which is called the moduli space of classical solution. Concretely
speaking,  if a space-time dependent solution to the classical
equation of motion cannot manifest a certain symmetry of the field
theory explicitly, then there must have some parameter in the
solutions to specify the symmetry. The number of independent
parameters in the solution is the dimension of the
 moduli space of the classical solution. At the late stage of 1970s, both
theoretical physicists and mathematicians employed various distinct
methods to determine the dimension of instanton moduli space.  A.S.
Schwartz \cite{schw}, Michael Atiyah, Nigel Hitchin and Isadore
Singer \cite{atiya1} used the celebrated Atiyah-Singer index theorem
in algebraic geometry to have identified the dimension of $SU(2)$
instanton moduli space as $8|k|-3$. At the same time, Jackiw and
Rebbi \cite{jackiw3}, Lowell Brown, Robert Carlitz and Choon-Kyu Lee
\cite{bclee} determined the dimension by  analyzing directly degrees of
freedom in the instanton solution and calculating the number of
fermionic zero-modes of the Dirac operator in the instanton
background, respectively. Moreover, the number of independent
parameters for the instanton solution to the Yang-Mills theory with
a general gauge group was worked out by theoretical physicists
Claude Bernard, Norman Christ, Alan Guth and Erick Weinberg
\cite{guth} as well as mathematician Atiyah, Hitchin and Singer
\cite{atiya2}. Further, mathematicians \cite{ward} solved the
problem how to construct an instanton solution for an arbitrarily
given Chern number by using Roger Penrose's twistor description
\cite{penrose} to Yang-Mills theory. This construction showed the
power of algebraic geometry in gauge theory \cite{ward}.
Finally, mathematicians proved that a finite action solution to
the Yang-Mills theory on a compactified Euclidean space in four
dimensions  must be an instanton solution \cite{simons}. At the
end of 1970s, all the puzzles on instanton solution and the
instanton moduli space had been cleared. To summarize, the joint
efforts made by both physicists and mathematicians at the late of
1970s had laid the foundation for the breakthrough made in 1980s
in understanding differential
topological structure of a simply-connected smooth four-manifold.\\

On the other hand, in the same time theoretical physicists started
  investigating physical effects induced by instanton. In the
middle of 1970s, Jackiw and Rebbi \cite{jackiw1}, Curtis Callan,
Roger Dashen and David Gross \cite{gross1} found the vacuum
structure of a non-Abelian gauge theory is highly nontrivial: the
Yang-Mills theory with gauge group $SU(2)$ has  vacua with an
infinite number of degeneracies. These vacua are classified into
distinct homotopy classes  by the mapping from $S^3$ to $SU(2)$ and
 characterized by topological indices.  People usually thought
these topological vacua should be absolutely stable since
topological numbers should prevent the vacuum from decaying.
However, the existence of instanton breaks this naive physical
pattern. Gerard 't Hooft first studied quantum gauge theory in the
instanton background \cite{hooft1}. He found that an instanton can
cause the transition  between two topological vacua if the
difference of their topological indices equals to the Chern number
carried by the instanton. This is the famous tunneling effects produced by
instanton in a quantum gauge theory. A clear physical interpretation
on the tunneling effect in quantum chromodynamics was further given
by Callan, Dashen and Gross \cite{gross2}. The tunneling phenomenon
has lifted the degeneracy of topological vacua and the true vacuum
state is the so-called $\theta$-vacuum, the superposition of
topological vacua. However, if the theory has massless fermions,
then the tunneling effect produced by instanton disappears. The
reason for this phenomenon is that a massless fermion carries not
only usual fermionic charge (electric charge, lepton number or
baryon number, depending on physical objects fermionic fields
represent), but also a fermionic charge which changes sign under
mirror (parity) transformation. In physics, it is called that the
massless fermionic field theory has  chiral symmetry (or equivalently
axial vector $U_A(1)$ symmetry). However, in the presence of
instanton configuration, this symmetry is violated by quantum
correction. Thus the fermionic charge that flips a sign under the mirror
reflection transformation  is not conserved and acquires a
contribution proportional to the instanton number, which is induced
by quantum correction. This phenomenon in physics is called that
$U_A(1)$ symmetry suffers from chiral anomaly \cite{anomaly}. At the
late stage of 1970s, it was realized that chiral anomaly is
independent of  perturbative calculation of quantum field theory and
has a topological origin. The violation of the axial fermionic
charge is equal to the difference of the left- and right-handed
fermionic zero modes of the Dirac operator in the instanton
background \cite{indano}. This means that the Dirac operator acting
on the massless fermionic fields in the instanton background must
have non-paired zero modes. Since a fermionic field is represented
by a Grassmann quantity, so the existence of non-paired zero modes
leads to vanishing integration over fermionic fields and hence the
tunneling effect is suppressed by massless fermions. ' t Hooft used
the integration property of the Grassmann quantity to have realized
that this phenomenon is actually a topological section rule for the
physical process. Some gauge invariant quantities carrying the
fermonic charges which change sign under parity transformation can
absorb the fermionic zero modes and present non-vanishing
expectation values, and hence contribute to the 't Hooft quantum
effective action. 't Hooft used this idea to have solved the
notorious $U_A(1)$ problem in particle
physics.\\

The instanton background  can also cause zero modes for the
operators acting on bosonic fields such as scalar and vector fields.
However, the origins of these bosonic  zero modes are completely
different from the fermionic ones for the Dirac operator. As
mentioned before, the instanton solution must contain some
parameters to manifest gauge and space-time symmetries of the
theory. The variations of these parameters do not alter the action
or potential energy of the theory. Therefore, the bosonic operators
have zero modes along the directions represented by these
parameters. In physics one can consider these bosonic modes as the
Goldstone modes corresponding to the breaking of certain global
symmetries in the directions represented by the parameters. Viewed
from the geometry of instanton moduli space, these bosonic zero
modes are actually tangent vectors to the moduli space and hence the
number of these zero modes is the dimension of the instanton moduli
space, since the dimension of a tangent space to a manifold is equal
to the dimension of the manifold.  One usually  makes use of a
so-called collective coordinate method  to handle the bosonic zero
modes in quantum field theory. In the path integral description of
quantum field theory, the essence of collective coordinate  method
is separating the integrations over zero modes from those non-zero
modes. After the non-zero modes have been integrated out, the path
integration over the bosonic fields reduces to an integration on the
instanton moduli space and the key point in this process is how to
define the integration measure on the instanton moduli space. Note
that this is actually the physical idea used by Witten to describe
the Donaldson invariant with a quantum gauge theory. It should
emphasize that the above mentioned 't Hooft's instanton calculus is
a pioneer work in non-perturbative calculation. During these past
thirty years, most of important developments in non-perturbative
quantum field theory such as instanton calculus in supersymmetric
gauge theory \cite{ins1,ins2,ins3,ins4}, topological quantum field
theory \cite{witt1} and the Seiberg-Witten duality \cite{seiwit} are
somehow the prolongs of 't Hooft's instanton calculus.  't Hooft
converted the complicated calculation in the instanton background
into a central potential problem in quantum mechanics, and it is
still not an easy task to repeat his calculations  for a beginner
despite that
 more than thirty years have passed.

\section{Supersymmetry, Supersymmetric Gauge Theory, $R$-symmetry
and Super-instanton Calculus}
\label{sect4}
\renewcommand{\theequation}{4.\arabic{equation}}

Supersymmetry plays a vital role in constructing topological
Yang-Mills theory. It entered physics in the middle of 1970s and
brought about a relativistic quantum field theory with supersymmetry
\cite{wess}. It is well known that  all the elementary  particles
are classified into two types, bosons with integer spin and fermions
of half-integral spin, and they are described by Lorentz tensors
(scalar, vector etc.) and spinors, respectively. It
should be emphasized that supersymmetry
 is a fermionic-type space-time symmetry, i.e., its generator is a
conservative charge with spin $1/2$. Therefore, supersymmetry can
collect bosons and fermions into one multiplet. The supersymmetry
transformation turns a boson into a fermion and \emph{vice versa}.
Supersymmetry is the only generalization of the Poincar\'{e} group
permitted by physical requirements. All the bosonic symmetries other
than those in the Poincar\'{e} group impose too much restrictions on
a relativistic quantum field theory \cite{colman2} so that the
scattering amplitude of interacting particles contradicts with
experimental observation.\\

Just like the Lorentz symmetry is an isometry symmetry of the
Minkowski space, with regard to supersymmetry one can also introduce
fermionic type coordinates to define a superspace. Supersymmetry and
the Poincar\'{e} symmetry then constitute an isometric symmetry of
superspace \cite{salam}.  For the simple $N=1$ supersymmery, the
bosonic and fermionic fields in one supermultiplet form a superfield
defined in the superspace with the isometry group $OSp(1,3|4)$. The
superfields are classified into chiral or vector superfield,
depending on the field content sharing a supermultiplet.  The
classical action of a supersymmetric field theory in superspace can
always divide into two terms: the K\"{a}hler potential and
superpotential. It should emphasize that the superpotential has an
elegant feature: it is an analytical (or anti-analytical) functional
of chiral superfields, and this feature is called holomorphy (or
anti-holomorphy). Since fermionic fields are represented by the
Grassmann variables, so bosonic and fermionic fields in one
supermultiplet  contribute opposite quantum corrections. As a
consequence of supersymmetric Ward identities, the perturbative
quantum correction can be partially or fully canceled. This fact
results in the nonrenormalization theorem, which states the
superpotential in a supersymmetric quantum field theory receives no
quantum correction. If formulated in superfield, this theorem means
that the superpotential at quantum level keeps its holomorphic form.
Therefore, a supersymmetric field is much more easily solvable than
a non-supersymmetric theory.\\

A supersymmetric field theory has a rich mathematical structure.
According to supersymmetry algebra, two consecutive supersymmetry
transformations lead to a space-time coordinate translation, thus
the generators of  a supersymmetric transformation  is directly
related to the Hamiltonian of the theory. This implies that the
ground state of a supersymmetric theory must be a zero-energy state.
In particular, if a supersymmetric theory has no spontaneous
breaking, its zero-energy vacuum state must exist. Therefore, the
supersymmetry generator is a nilpotent operator when acting on a
ground state of a supersymmetric field theory. This feature is
similar to the exterior differential operator acting on differential
forms in differential geometry, and consequently, the Hamiltonian of
the theory corresponds to the Laplacian operator acting the harmonic
differential forms. Therefore, if we can can construct a special
supersymmetric quantum field theory whose Hilbert space consists
only of the vacuum states of the theory, then its physical states
are homological class of the supercharge. Further, if a
correspondence between the homological class of the supercharge and
the homology (or cohomology)  class of the space-time manifold on
which the theory is defined can be established, one can use a
supersymmetric quantum field theory to study differential topology
of a smooth manifold.\\

We should specify $R$-symmetry in a supersymmetric gauge theory
since it plays a crucial role in reproducing the Donaldson
polynomial invariants in terms of topological Yang-Mills theory. The
generator of $R$-symmetry and the supercharge together with those generators
for the Poincar\'{e} symmetry constitute the whole supersymmetry
algebra. $R$-symmetry is an internal-like symmetry, so its
generator(s) has (have) only non-trivial commutation relations with
supersymmetry generators. It is a chiral symmetry and implements the
automorphic chiral rotations among supercharges if they are
represented by the Weyl spinors (or axial symmetry if the
supersymmetry generators represented by four-component Majorana
spinors). For an $N$-extended supersymmetry, if the supersymmetry
algebra has central extension, the $R$-symmetry group is usually
$U(N)$; While in the presence of central charge, it is $USp(N)$, the
compact version of the symplectic group $Sp(N)$. Accompanying the
supersymmetry, the $R$-symmetry has a natural representation in a
supersymmetric field theory. The field functions in a supermultiplet
carry the representations of $R$-symmetry  according to commutative
relations between the supercharge and the generator of $R$-symmetry.
For $N=1,2$ supersymmetric Yang-Mills theories, their $R$-symmetries
are $U_R(1)$ and $U(2)=U_R(1)\times SU(2)$, respectively, and each
step of  supersymmetry transformation on a field function in a
supermultiplet increases its $U_R(1)$-charge by one. The
$R$-symmetry of $N=4$ supersymmetric Yang-Mills theory is described
by the non-Abelian group $SU(4)$. In a classical supersymmetric
gauge theory, according to the supersymmetry algebra, the
$R$-symmetry current, the energy-momentum tensor and the
supersymmetry current constitute a superconformal current
supermultiplet. At quantum level, the $U_R(1)$ symmetry usually
becomes anomalous due to its chiral feature. Especially, the chiral
anomaly of $U_R(1)$ current, the trace anomaly of the
energy-momentum tensor and the gamma-trace anomaly of the
supersymmetry current share a superconformal anomaly supermultiplet,
and the common anomaly coefficient is proportional to the beta
function of supersymmetric Yang-Mills theory. This feature is very
useful for determining the perturbative part of quantum effective
action for the $N=2$ supersymmetric Yang-Mills theory since its
perturbation theory is one-loop exhausted. \\

In the remained part of this section, we shall introduce the
instanton calculus in a supersymmetric gauge theory. Witten actually
employed the supersymmetric instanton calculus to reproduce the
Donaldson invariants in terms of the observables of topological
Yang-Mills theory \cite{witt1}. In comparison with the instanton in
the usual Yang-Mills theory, a supersymmetric instanton have new
features. In the following we illustrate these features using an
$N=1$ supersymmetric gauge theory. First, the instanton has a
fermionic partner with definite chirality and it is automatically a
fermionc zero mode of the Dirac operator in the instanton background
\cite{ins2}. Moreover, the fermionic partner of (anti-)self-dual
instanton must be described a (right-)left-handed Weyl spinor.
Second, just like the classical Yang-Mills theory has a conformal
symmetry, a classical supersymmetric Yang-Mills theory has a
superconformal symmetry consisting of the usual conformal space-time
symmetry, Poincar\'{e} supersymmetry and conformal supersymmetry
\cite{ins2}. The conformal supersymmetry transformation parameters
are space-time coordinate dependent, and the conformal supersymmetry
transformation on the instanton solution yields the conformal
supersymmetric partner of the instanton. Naturally it is also a
fermionic zero mode of the Dirac operator in the instanton
background, but it has opposite chirality with the Poincar\'{e}
supersymmetric partner. The reason is that  the generators for the
Poincar\'{e} supersymmetry and conformal supesymmetry have opposite
chiralities.  Finally, it should be emphasized that the instanton
configuration breaks the conformal supersymmetry and preserves only
the Poincar\'{e} supersymmetry. This is the reason why there exist
the fermionic zero modes: it is just the action of conformal
supersymmetry generators on the instanton solution that leads to
the fermionic zero modes. In the supersymmetric instanton calculus,
there are two approaches to proceed. The first one is using the
original instanton calculus invented by 't Hooft and one deals with
the fermionic zero modes in the same way as manipulating the usual
fermionic zero modes in the instanton background, the only
difference is that the present fermionic zero modes are in the
adjoint representation of gauge group rather than in the fundamental
representation. The other way, which has been used in most of
literatures, is considering the instanton and the fermionic
zero-modes as an instanton supermultiplet, i.e., superinstanton
\cite{ins2}. In this way, the instanton moduli space of a
supersymmetric gauge theory is described by the parameters in the
instanton solution and the Grassmann parameters in the fermionic
zero modes, which reflect the breaking of conformal supersymmetry.
The super-instanton moduli space is thus a supermanifold. Therefore,
the integration measure in the instanton moduli space in a
supersymmetric gauge theory consists of not only the integration
over the parameters denoting the center and the size of instanton,
but also the integration over the Grassmann parameters representing
the conformal supersymmetry in the fermionic zero modes. Because the
fermionic zero modes have definite chiralities, they carry the
$U(1)$ $R$-symmetry charges. So the moduli space of super-instanton
can be thought as a ``differential form" fibre bundle space with the
usual Yang-Mills instanton moduli space as the base manifold, and
the degrees of "differential forms" are just R-charges. This is the
key idea that Witten used  to reproduce the
Donadlson invariants in terms of topological Yang-Mills theory.\\

For a supersymmetric gauge theory with extended supersymmetries,
i.e., $N=2$ or $N=4$ case, the theory has scalar fields, which are
components of the extended supermultiplet \cite{ins4}.  This makes
the instanton calculus delicate.  The reason is that the classical
action of an extended supersymmetric gauge theory contains
additional scalar potential and the Yukawa coupling term among
scalar and fermionic fields. So if we have a careful look at the
classical equation of motion, it seems that the fermionic and scalar
zero modes arising at the lowest order of gauge coupling become
non-zero modes. In this case the instanton solution is called a
quasi-instanton and the parameters defining the instanton moduli
space are called quasi-collective coordinates.  However, as long as
the theory has chiral $U(1)$ R-symmetry like $N=1$, $2$
supresymmetric Yang-Mills theories, it must suffer from chiral
anomaly. Then according to the topological origin of chiral anomaly
described by the Atiyah-Singer index theorem in algebraic geometry,
there must exist fermionic zero modes for the Dirac operator and the
scalar field zero modes for the Laplacian operator in the instanton
background.  But $N=4$ supersymmetric Yang-Mills theory has only
$SU(4)$ non-Abelian chiral R-symmetry, the situation is different.
Another problem is that the scalar potential causes spontaneous
breaking of gauge symmetry, i.e., the scalar field has non-vanishing
vacuum expectation value. Rigorously speaking,  the theory has no
instanton solution in this case, and we must use the notion of
constrained instanton proposed by Ian Affleck \cite{aff}. This kind
of instanton behaves similarly as the Yang-Mills instanton at
short-distance (or equivalently, the vacuum expectation value of
scalar field is very small), and presents the exponential decay at
long distance. This is actually a good phenomenon for instanton
calculus since the infrared divergence caused by the large size
instantons can be cured. Witten used the instanton calculus in a
twisted $N=2$ supersymmetric $SU(2)$ Yang-Mills theory to reproduce
the Donaldosn invariant, where it does not matter whether the scalar
field has vacuum expectation value or not. As it will be introduced
later, if the $SU(2)$ gauge symmetry breaks spontaneously to $U(1)$,
the Donaldson invariants can be calculated from the magnetic dual
theory of $N=2$ supersymmetric Yang-Mills theory and the problem
even becomes much more simple \cite{witt5}.

\section{Differential Structure of
Four-dimensional Manifold and Mathematical Construction of
Donaldson Invariant}

It was one of the most perplexed and subtle problems in differential
geometry to distinguish differential structures of a differentiable
manifold. Two manifolds are called homeomorphic if there exists a
continuous one-to-one mapping between them. All the manifolds with
same dimensions can be classified into equivalent classes according
to the homeomorphic mapping. Further, if the homeomorphic mapping
between two differential manifolds is differentiable, then these two
manifolds are said to be diffeomorphic. This means that the
manifolds sharing the same homeomorphic class can be more
elaborately distinguished in terms of the diffeomorphism mapping.
The manifolds presenting the same topology may have distinct
differential structures. Roughly speaking, the notion of
differential structure on a manifold describes how a coordinate
atlas on the manifold is assigned and how the open coverings of the
manifold are ``glued" together. It reflects the smoothness of a
differential manifold. The notorious examples of illustrating
differential structure are seven-dimensional spheres: $S^7$ and the
Milnor exotic spheres $\widetilde{S}^7$ are homeomorphic but not
diffeomorphic. The $\widetilde{S}^7$ can be created from $S^7$ by
cutting it along the equator, transforming the boundary of one
hemisphere and then ``gluing" the two halves back. Mathematicians
proved that low-dimensional manifolds (dimension less than 4) have
unique differential structure, but higher dimensional manifolds
usually have a variety of differential structures, and the most
delicate case is the four-dimensional manifold,
the physical space-time.\\

 The Donaldson polynomial
invariants distinguish differential topological structure of a
simple-connected smooth manifold in four dimensions \cite{dona}.
Before its invention, the invariant characterizing the topological
structure of a four-dimensional differential manifold  is the
characteristic class on the cotangent bundle of the manifold
constructed by the Russian mathematician Lev Semenovich Pontrjagin.
This characteristic class is invariant under the homeomorphism
transformation on the manifold, but the topological number obtained
from the integration of the Pontrjagin class over the manifold can
only describe  the topology of a differentiable manifold and has little
power to distinguish the delicate differential topological
structure. At the early stage of 1980s, a mathematician Michael
Freedman realized that the differential topology of a simply
connected smooth four-dimensional manifold $M$ is related to the
intersection form defined on the second integer cohomology class
$H^2(M,Z)$ over $M$ \cite{free}, which is a generalization of the
Pontrjagin class on a four-dimensional manifold. In mathematical
terminology, an intersection form $\Omega_M(\alpha,\beta)$ is a
mapping from the $H^2(M,Z)$ space to an integer set $Z$. It is
defined by taking the wedge (or exterior) product $\alpha\wedge
\beta$ of two elements $\alpha$, $\beta$ in $H^2(M,Z)$ and
integrating over the top homology class $H_4(M,Z)$ on the manifold
$M$. If  a set of orthonormal basis in the $H^2(M,Z)$ space is
chosen, then the intersection form can be written as a $n\times n$
matrix of integer-valued element, where $n$ is the dimension of
$H^2(M,Z)$ and it is an even number. Further, since the structure
group of the cotangent bundle on a four-dimensional orientable
manifold is $SO(4)$ and the fourth rank antisymmetric tensor
$\epsilon_{\mu\nu\lambda\rho}$ is an $SO(4)$ invariant,  all the
elements in $H^2(M,Z)$ can be classified into the orthogonal
self-dual and anti-self-dual sectors through a projection of
$\epsilon_{\mu\nu\lambda\rho}$ on the second-rank antisymmetric
tensors. Consequently, the intersection matrix decomposes two
$n/2\times n/2$ blocks and they are the matrix representation of an
intersection form realized on the self- and anti-self-dual basis of
$H^2(M,Z)$. The reason why an intersection form can describe the
differential topological structure of a smooth four-manifold is that
the homeomorphism transformation induced a homeomorphism
transformation on $H^2(M,Z)$ and correspondingly, the intersection
matrix undergoes a similar transformation. As a consequence, the
intersection matrices are classified into equivalent classes. In
this way, the equivalent classes of the intersection forms  are
related to the classification of differential topology of the
manifold. At the beginning of 1980s, Freedman proved that depending
on whether the diagonal elements of an intersection matrix are  even
or odd numbers, the classification of the simply-connected smooth
four-manifolds under the homeomorphism can almost be uniquely
determined by the equivalent classes of the intersection matrices
\cite{free}. In 1983, Simon Donaldson proved a crucial theorem that
a positive-(or negative) definite intersection matrix of  a
simply-connected smooth four-manifold can always be converted into a
positive (or negative) unit matrix \cite{dona}. This theorem has
provided a criteria for the smoothness (infinite differentiability)
of a simply  connected four-manifold.  In proving this theorem,
Donaldson used the notion of instanton moduli space in the Yang-Mills
theory, and the number of the eigenvalues equal to $+1$ in an
intersection matrix is actually the number of singularities on the
instanton moduli space. Further, Donaldson defined  a mapping from
the second integer homology group $H_2(M,Z)$ on a simply connected
smooth four-manifold $M$  to the second cohomology group $H^2({\cal
M}_k)$ over the instnaton moduli space ${\cal M}_k$. This is
actually a ``differential form " on the Yang-Mills instanton moduli
space defined by the Donaldson mapping. Then Donaldson took the
wedge product of $d$ differential forms defined on $H_2(M,Z)$ but
taking values in $H^2({\cal M}_k)$ and performed integration over
the compactified instanton moduli space ${\cal M}^c_k$. In this way,
he finally constructed the powerful integer-valued polynomial
invariants of order $d$ to distinguish the delicate differential
topological structure of a simply-connected smooth four-manifold.

\section{Donaldson Invariant as Physical Observable
of Witten's Topological Yang-Mills Theory}

The differential topological invariants constructed by Donaldson
using the Yang-Mills instanton moduli space  had been  re-derived
from a relativistic quantum field theory by Witten in a physical way
\cite{witt1}. In the following we shall explain the physical ideas
behind this field theory reproduction. First, as a topological
invariants depicting differential topology of a manifold, it must be
independent of the metric of the manifold. In physics, the Einstein
equation tells that the energy-momentum tensor of matter field is
the source of resulting in the variation of the metric on space-time
manifold. So if the physical observable of a relativistic quantum
field theory has nothing to do with the metric of the space-time
manifold on which the theory is defined, the vacuum expectation
value of energy-momentum tensor must equal to zero. Further,
the Hamiltonian of the field theory is a component of the
energy-momentum tensor and moreover, by definition, the zero mode of
the Hamiltonian operator is a vacuum state of quantum field theory,
thus the local quantum states in the Hilbert space of the anticipated
theory should be only the vacuum  states (or equivalently the
contribution to the
 physical observable from local quantum excited states should be
 suppressed). This fact reminds us that the anticipated quantum field
 theory should have supersymmetry or a certain symmetry similar to
 supersymmetry since it can make the quantum corrections coming from
 local bosonic and fermionic excited states canceled. In addition, as
 discussed before, the notion of instanton
 moduli space and the cohomology group on it had been used to
construct the Donaldson invariant. From the introduction to the
super-instanton moduli space in a supersymmetric gauge theory, we
know that the instanton moduli space of $N=2$ supersymmetric
Yang-Mills theory  is a ``differential form" fibre bundle over the
usual Yang-Mills instanton moduli space and the degree of ``the
differential form" is the chiral $U_R(1)$-charge. However, one
cannot use the standard $N=2$ supersymmetric Yang-Mills theory to
get the Donaldson invariant. The physical reason is that $N=2$
supersymmetry is not powerful enough to eliminate all the quantum
corrections contributed from local quantum excited states. The other
reason comes from mathematical side.  As we know from the knowledge
in algebraic topology, the homology group of a manifold
straightforwardly describes its topological structure. A geometrical
or physical object must have a one-to-one correspondence with an
element in the homology group to reflect the topology of the
manifold. For example, the reason why the de Rahm cohomology group
can describe the topology of a certain manifold is because it
establishes a correspondence with the homology group through the
Stokes theorem. Therefore, to construct a quantum field theory
description to the Donaldson invariant, the generator of the
supersymmetry-like symmetry  must be nilpotent like the boundary
operator (or exterior differential operator) and the physical states
must form a homology class with respect to the symmetry generator.
The nilpotency of the symmetry generator provides a possibility that
establishes a correspondence between a physical observable of
quantum field theory and an element in the homology (or cohomology)
group on a differential manifold. As mentioned before, the
supersymmetry generator is nilpotent only when acting on vacuum
states. The problem is how to achieve its nilpotency and in the meantime to
generate the symmetry of the full
theory.\\

In fact, in a quantum gauge theory, we have such a symmetry at hand.
As it is well known, a gauge field theory has redundant degrees of
freedom due to gauge symmetry, so when performing path integral
quantization of a non-Abelian gauge field theory, one must first fix
the gauge  to make the path integral measure well defined. This
problem was beautifully solved by Russian mathematical physicists Ludvig
Faddeev and Victor Popov \cite{ghost}. Consequently, the explicitly
gauge-invariant action is replaced by an effective action containing
ghost fields and the local gauge symmetry manifests itself as the
BRST symmetry named after its discovers C. Becchi, A. Rouet, R.
Stora \cite{brs} and I.V. Tyupin \cite{tyu}. The generator of this symmetry is
nilpotent and it is a fermionic-like quantity. The ghost fields are
non-physical field functions since  they are scalars with respect
to space-time symmetry but obey fermionic statistics. In addition,
the gauge-fixed effective action also has a $U(1)$ ghost number
symmetry, which are charged only with ghost fields. The usual
convention is that the BRST charge should carry a ghost number $-1$
and then each BRST transformation on a field function increases the
ghost number by one. It turned out that the BRST transformation and
the ghost field also admit elegant geometric descriptions. As we
know, the classical action of a gauge field theory is a functional
on the gauge field configuration space ${\cal U}=\{A\}$, which is an
infinite dimensional affine space with trivial topology. However,
due to the gauge symmetry, the physical field configuration space is
actually the infinite dimensional gauge orbit space ${\cal U}/{\cal
G}$, here ${\cal G}=\{g(x)\}$ is the gauge transformation group,
which consists of all of the gauge transformations $g(x)$ and is an
infinite dimensional group due to the space-time coordinate
dependence of the group element
\cite{atijones,atisi,wuzee2,reina1,houzhang}. Further, ${\cal U}$
can be naturally considered as a fiber bundle space with the base
manifold ${\cal U}/{\cal G}$ and the fibre ${\cal G}$. Each point
$a(x)=a_\mu dx^\mu$ on ${\cal U}/{\cal G}$ is a 1-form connection on
the principle $G$-bundle $P(M,G)$ over the base manifold $M$ and
each point $A(x)$ on ${\cal U}({\cal U}/{\cal G}, {\cal G})$ just
differs a gauge transformation with $a(x)$ implemented by an element
of ${\cal G}$, i.e., $A=g^{-1}a g+g^{-1}dg$. Like $P(M,G)$, one can
introduce a connection ${\cal A}$ on ${\cal U}({\cal U}/{\cal G},
{\cal G})$ and use it to decompose a cotangent vector
$\widetilde{\delta}A$ at the point $A\in {\cal U}$ into vertical and
horizontal components. Then the ghost field is the `` Maurer-Cartan
form " $v(x)\equiv g^{-1}(x)\delta g(x)$ in the connection ${\cal
A}$, and the BRST transformation $\delta A$ on the gauge field
$A(x)$ is the variation of the cotangent vector
$\widetilde{\delta}A$ along a fibre of ${\cal G}$
\cite{atijones,atisi,wuzee2,reina1,houzhang}. Naturally the BRST
transformation preserves $a(x)$ but produces a infinitesimal gauge
variation on $A(x)$: $\delta a=0$ and $\delta A=Dv$, where $D$
denotes the gauge covariant derivative operator defined
with respect to $A(x)$.\\

The above geometric descriptions to the BRST transformation and  the
ghost field are quite mathematical. We prefer to a more physical way
and re-interpret it in an extended principle fibre bundle theory
\cite{mieg, bono}. First, a principle $G$-bundle $P(M,G)$ over the
base $M$ is a fibre bundle whose fibre space and the structure group
are identical, both being a Lie group manifold $G$. The connection
one-form on the principle $G$-bundle in the cotangent space approach
is $\omega (x)=g^{-1}(x)A(x) g(x)$$+g^{-1}(x)dg(x)$ \cite{eguchi}
and now $g=g(x)$ is an element of Lie group fibre located at the
point $x\in M$. The projection of the first term $ g^{-1}Ag$ on the
base manifold $M$ yields the gauge field $A=A_\mu^a(x) T^a dx^\mu$
and the second one $g^{-1}dg=\phi^a (x) T^a $ is the vertical
Maurer-Cartan form with one-form $\phi^a$ satisfying the
Maurer-Cartan structure equation.
Now we extend the base manifold $M$ to $M\times G$ and define an
extended principle $G$-bundle ${\cal P}$ over the base manifold
$M\times G$ with the same Lie group fibre $G$, then the coordinate
of the fibre space space $G$ becomes $g=g(x,\theta)$. Near the unit
element it can be locally written as $g(x,\theta)=\exp [\theta^a (x)
T^a]$, where $\theta^a$ ($a=1,2,\cdots, \dim G$) are the parameters
of Lie group $G$. As a consequence, the connection one-form on the
extended principle $G$-bundle ${\cal P}$  reads $\widehat{\omega}
(x,\theta)=g^{-1}(x,\theta) A(x) g (x,\theta) + g^{-1}(x,\theta)
(d+\delta) g (x,\theta) \equiv\omega (x,\theta)+ v(x)$ and
$\delta\equiv d\theta^a \otimes\partial/\partial \theta^a$ is an
exterior differential operator on the Lie group manifold. Then the
ghost field $c^a(x)$ in the gauge-fixed Yang-Mills theory
 plays precisely the role of the Maurer-Cartan one-form in the group
 space, $v(x)=g^{-1}(x,\theta)\delta g(x,\theta)\equiv c^a(x)T^a$,  and the
BRST transformation is the exterior differentiation (i.e., the
action of $\delta$) on the Lie group manifold
\cite{mieg,bono,stora2,zumi2}. Since the geometric counterpart of
the BRST charge is the exterior differential operator $\delta$ and
the ghost fields are sections of a certain exterior algebra bundle,
the geometric description  thus explains why the BRST charge is
nilpotent and the ghost fields present the anticommuting character
despite that they are scalar fields with respect to the space-time
symmetry. Later we shall show that the geometric description to the
BRST transformation plays an essential role in deriving the
celebrated Stora-Zumino descent equations
\cite{bono,stora2,zumi2,zumi}, which is another key element used by
Witten to construct the physical observables for reproducing the
Donaldson polynomial invariants in
terms of quantum field theory. \\

In physics, based on the BRST symmetry, physicists developed a
canonical quantization method starting from the gauge-fixed
effective action \cite{japan}, and it is called the BRST
quantization, in which all the physical states belong to holomogical
classes of the BRST charge, i.e., they are BRST invariant modulo
those BRST trivial
states which have zero norms.\\

 According to the above gathered  materials, the anticipated quantum field
 theory should be a Yang-Mills gauge theory with a global symmetry
 presenting the features of both supersymmetry and BRST symmetry.  In
 1988 Witten  constructed such a theory by twisting $N=2$
 supersymmetric Euclidean Yang-Mills theory, and this theory is now called the
 topological Yang-Mills theory \cite{witt1}. The twisting operation  means
 choosing a $SU(2)_X$ subgroup of the space-time Euclidean Lorentz group
 $SO(4)\cong SU(2)_X\times SU(2)_Y$
and another $SU(2)_R$ subgroup from the $R$-symmetry group $U(2)=
SU(2)_R \times U(1)_R$ and taking the diagonal $SU(2)$ subgroup of
the direct product $SU(2)_X\times SU(2)_R$ (i.e., decomposing the
representation of direct product group $SU(2)_X\times SU(2)_R$ in
terms of the representation of its subgroup group $SU(2)$), and then
using this $SU(2)$ group and the remained group $SU(2)_Y$ and
$U(1)_R$ to form a new space-time symmetry group $SU(2)\times
SU(2)\times U(1)\cong  SO(4)\times U(1)$. Consequently, all the
field functions and  supercharges of $N=2$ supersymmetric Yang-Mills
theory are re-classified according to new space-time symmetry group
$SU(2)\times SU(2)\times U(1)$ and they become twisted field
variables and symmetry generators. Then we rewrite the classical
action of $N=2$ supersymmetric Yang-Mills theory in terms of twisted
field functions and the twisted theory presents a BRST symmetry. The
BRST charge comes from the twisting of the right-handed supercharge
carrying $R$-charge, and it is a scalar with respect to the new
space-time symmetry group $SO(4)$ and carries R-charge $-1$
inherited from the $N=2$ supersymmetric Yang-Mills theory before the
twisting. In comparison with the counterpart in the BRST
quantization, the $U(1)$ $R$-symmetry is the ghost number symmetry
and the R-charge is identical to the ghost number. It is a miracle
that both the twisted classical action and the energy-momentum
tensor is BRST trivial as expected. With the idea of BRST
quantization of gauge theory, the theory has no local quantum
excited states and its physical observables constitute a homology
class of the BRST charge.
The physical process described by the theory is only the transition
among topological vacua caused by instanton tunneling effect. The
transition amplitudes can be reduced to integrals over the instanton
moduli space since all the quantum corrections contributed from
local excited states cancel, attributing to the powerful BRST
symmetry \cite{witt1}.\\

Now the confronting problem is how to construct appropriate physical
observables in the topological Yang-Mills theory to reproduce the
Donaldson polynomial invariants. According to the mathematical
construction on the Donaldson polynomial invariants introduced
before, the to-be-constructed physical observables, in addition that
they are gauge invariant, should satisfy two requirements. First, as
emphasized previously, the expected physical observable must have a
correspondence with an element in a homology (or cohomology) class
on the space-time manifold so that it can indeed represent a certain
topological invariant. Second, the Donaldson polynomial invariant
was constructed mathematically with a mapping from the integer
homology group on a differential manifold in four dimensions to the
cohomological group on the instanton moduli space of Yang-Mills
theory, thus the anticipated physical observable must also reflect
the correspondence between the cohomological class over the
instanton moduli space and the homological (or cohomological) class
on the space-time manifold. On the other hand, in topological
Yang-Mills theory the field functions are graded by ghost number
 (or $R$-charge called in the untwisted $N=2$ supersymmetric
Yang-Mills theory), and the BRST transformation on a field function
increases the ghost number by one, just behaving as an exterior
differential operator acting on a differential form. Based on these
analogues, Witten ingeniously made use of the descent equations in
 an anomalous gauge theory and successfully fulfilled the above two
 geometrical
requirements \cite{witt1}.\\

The chain of descent equations plays crucial role in attaining the
physical observables for the Donaldson polynomial invariants. It was
found by Raymond Stora,  Bruno Zumino and Juan Ma\~{n}es
\cite{stora2,zumi2,zumi}, Bonora and Cotta-Ramusino \cite{bono} in
evaluating the non-Abelian chiral anomaly through observing the
Wess-Zumino consistency condition \cite{wezu2}. One chain of descent
equations is precisely the local version of the Wess-Zumino
consistency condition \cite{wezu2} and hence provide a physically
legitimate basis of using solely differential geometry to find the
non-Abelian chiral anomaly \cite{zuwze} (or the Bardeen anomaly in
the covariant case \cite{bard}). In the following we use a physical
and geometrical hybrid way to introduce the descent equations.  In
an anomalous gauge theory consisting of chiral fermions coupled with
an external chiral gauge field $A_\mu=A_\mu^a T^a$, the anomalous
Ward identity can be formally expressed as the form that the anomaly
$G^a[A]$ comes from the action of gauge transformation generator
$X^a$ on the quantum effective action $W[A]$, which is is a
functional of the external gauge field $A_\mu$ obtained by
integrating out the chiral fermions in the path integral
quantization. Since the gauge transformation generator $X^a$ is a
local functional differential operator representation to the Lie
algebra generator $T^a$  of the gauge group, Julius Wess and Bruno
Zumino realized that there should naturally imposes a consistent
condition on the non-Abelian chiral anomaly $G^a[A]$. Reversely, the
non-trivial solution to this consistent condition should yield the
chiral anomaly $G^a[A]$. If one defines a top-rank differential form
$Q_{2n}^1\equiv c^a (x)G^a[A(x)]d^{2n}x$ with the ghost number one
on a $2n$-dimensional space-time manifold $M^{2n}$ by multiplying
the ghost field $c^a(x)$ with the non-Abelian anomaly $G^a[A(x)]$,
the Wess-Zumino consistent condition can elegantly expressed as
$\displaystyle \delta \int_{M^{2n}}Q_{2n}^1=0$, i.e., the BRST
invariance of the integration of the top-rank differential form
$Q_{2n}^1$ over the space-time manifold. This means locally the BRST
transformation of the above top-rank differential form with ghost
number one can be written as the exterior differentiation of a
differential form with one degree less
and the ghost number two, i.e., $\delta Q_{2n}^1 =-dQ_{2n-1}^2$.\\

It turned out that the above local version of the Wess-Zumino
condition is only one chain of the descent equations and they can be
derived from the geometric description  to the BRST transformation
and the ghost field, described in either the connection bundle
${\cal U}({\cal U}/{\cal G}, {\cal G})$ or the extended principle
$G$-bundle ${\cal P}(M\times G, G)$ introduced before
\cite{atijones,atisi}. The context of the first option is
mathematically rigorous but complicated
\cite{atijones,atisi,houzhang}. One must define a direct product
bundle $P(M,G)\times {\cal U}({\cal U}/{\cal G}, {\cal G})$, and
then introduce the total connection $A+{\cal A}$ and the
corresponding curvature, ${\cal F}=(d+\widetilde{\delta})(A+{\cal
A}) +(A+{\cal A})\wedge (A+{\cal A})$. It can be proved that ${\cal
F}$ has only non-vanishing horizonal lift in the product bundle
space (i.e, the lift along $M\times {\cal U}/{\cal G}$ direction).
Therefore, along the fibre of the connection bundle ${\cal U}({\cal
U}/{\cal G}, {\cal G})$, ${\cal F}=F$, the ``differential operator"
$\widetilde{\delta}$ on the connection space ${\cal U}$ reduces to
the BRST operator $\delta$ and the connection ${\cal A}$ becomes the
ghost field $v=g^{-1}\delta g$. Since the exterior differential
operators $d$, $\delta$ and $d+\delta$ are all nilpotent, we can
construct the Chern characteristic class $C_n ({\cal F})=
\mbox{Tr}({\cal F}^n)$ on $P\times {\cal U}$.
 Because the Chern class is
closed, and  hence according to the Poincar\'{e} lemma, it can be locally
written as an exact differential form. Further, there exists ${\cal
F}=F$ along the fiber of ${\cal U}$, so we have $\mbox{Tr}({\cal
F}^n)=(d+\delta) Q_{2n-1}(A+v)=\mbox{Tr}(F^n)=d Q_{2n-1}(A)$.
Expanding the above equation in powers of $v$ and comparing the
differential forms with
the same degrees, we  obtain  the chain of descent equations.\\

The descent equations  can also derived  from the extended principle
$G$-bundle ${\cal P}(M\times G, G)$ in a more easily accessible way
\cite{bono,stora2,zumi2}. First, we define the extended exterior
differential operator $\Delta\equiv d+\delta$ on the extended base
manifold $M\times G$. It is a formal direct sum of the exterior
differential operator $d$ on the space-time manifold and the
differential operator $\delta$ on the Lie group space, and is thus
nilpotent, $\Delta^2=0$; Second, we construct an extended Chern
characteristic class $C_n[\widehat{\Omega}
(x,\theta)]=\mbox{Tr}(\widehat{\Omega}^n)$ on the extended principle
$G$-bundle using the curvature
$\widehat{\Omega}(x,\theta)=\Delta\widehat{\omega}
(x,\theta)+\widehat{\omega} (x,\theta)\wedge\widehat{\omega}
(x,\theta)$. Third, there exists a natural result that  the
curvature of the extended principle $G$-bundle ${\cal P}(M\times G,
G)$ equals to the curvature of the original principle $G$-bundle
$P(M,G)$: $\widehat{\Omega} =\Delta
\widehat{\omega}+\widehat{\omega}\wedge\widehat{\omega}=
d\omega+\omega\wedge \omega =\Omega =g^{-1}(dA+A\wedge A)g=g^{-1}F
g$. Therefore, the extended Chern  class $C_n[\widehat{\Omega}]$
should equal to the Chern class $C_n[\Omega]$  and also the Chern
class $C_n[A]$ defined in terms of the pull-back $A$ of the
principle $G$-bundle curvature $\omega$ to the base manifold $M$:
$C_n[\widehat{\Omega}]=C_n[\Omega]$.
 Finally,  the closure and local exactness of the Chern class means
 $C_n[\Omega]=dQ_{2n-1}[\omega,\Omega]$. In the
local form $\Delta Q_{2n-1}[\widehat{\omega},\widehat{\Omega}]
=(d+\delta)Q_{2n-1}[\omega+v,\Omega] =dQ_{2n-1}[\omega,\Omega]$, we
expand the Chern-Simons form in powers of $v$:
$Q_{2n-1}[\omega+v,\Omega]=
\displaystyle\sum_{p=0}^{2n-1}Q^p_{2n-1-p}[v^p,\omega,\Omega]$.
Then comparing the differential forms of the same degrees on the
base manifold $M^{2n}$ and choosing $g=e$, the unit element of the
Lie group $G$, we immediately obtain the Stora-Zumino chain of
descent equations: $\delta
Q^{p}_{2n-1-p}[v^p,A,F]=-dQ_{2n-p-2}^{p+1}[v^{p+1},A,F]$ and $\delta
Q_0^{2n-1}=0$ ($0\leq
p\leq 2n-1$).\\

The chain of descent equations relates the physical
property of quantum field configuration to space-time geometry. It
shows the topological obstructions at various ranks for some
physical variables to have global definitions on  a space-time
manifold. It was invented by theoretical physicists and provided a
physical accessible tool of studying the global aspects of quantum
field theory  but described in terms of the  infinitesimal local gauge
transformation. This is the reason why it plays such a decisive role
in investigating physical phenomena originating from the
topologically non-trivial field configuration of a quantum gauge
theory. The chain of descent equations establishes a correspondence
between the BRST cohomology in field configuration space (or Hilbert
space) of quantum gauge theory and the de Rahm cohomology class over
the space-time manifold. Consequently, it relates the ghost number
directly to the degree of a certain differential form on the
space-time manifold. Thus it is no wonder that Witten naturally
employed an analogue of these equations to construct physical
observables to reproduce the Donaldson polynomial invariants in
topological Yang-Mills theory. Substituting the descent equations
into certain relevant Green functions, we can obtain various
anomalous Ward identities among the Green functions. The further
topological meaning embodied in the descent equations had been
tapped by Michel Dubois-Violette, Michel Talon and Claude-Michel
Viallet \cite{dvmv}, and it implies the existence of
$H^p_{2n-1-p}(\delta/d)$'s, the so-called $\delta$-cohomology modulo
$d$. This means those $Q^p_{2n-1-p}$'s entering the descent
equations must take away the terms of the form $Q^p_{2n-1-p}=\delta
\theta^{p-1}_{2n-p}+d\eta^{p-1}_{2n-p-2}$. The physical meaning that
$H^p_{2n-1-p}(\delta/d)$'s are various non-trivial anomalous terms
was clarified by Ma\~{n}es and Zumino \cite{manes}. For examples,
$Q_{2n-2}^1(v,A,F)$, the chain term to the order $v$ in the
expansion of the Chern-Simons form $Q_{2n-1}(A+v,F)$, describes the
consistent non-Abelian anomaly \cite{zuwze} and its $\delta$-trivial
form can be canceled by introducing local counterterm into the
quantum effective action \cite{manes}; $Q_{2n-3}^2$, the chain term
to the order $v^2$, represents the Schwinger term appearing in an
equal-time commutator of the Gauss law operators in the Yang-Mills
theory \cite{fadd2}; $Q^3_{2n-3}$, the chain term to the order
$v^3$, reflects the failure of the Bianchi identity  in the presence
of magnetic monopole \cite{jackwco,wuzee3}. The quantum anomalous
effects relevant to the chain terms of higher order $v$ have not
been found yet up to now.\\

 Since one chain of the descent equations is just the
local version of the Wess-Zumino consistency condition imposed on
the non-Abelian consistent anomaly, based on this observation,
Zumino, Yong-Shi Wu and Anthony Zee \cite{zuwze} invented the pure
differential geometrical approach to evaluate the non-Abelian chiral
anomaly with only simple calculations on the Feynman diagram to
determine the normalization constant, i.e., starting from the Chern
characteristic class and observing  its infinitesimal gauge
transformations, and then taking into account the Wess-Zumino
consistent condition to derive the anomaly. It is an elegant and
universal method, working not only for the non-Abelian chiral
anomaly in any even dimensions, but also for the pure gravitational
anomaly in $4k+2$ dimensions \cite{alga}. Especially, it is
extremely useful in calculating anomaly in higher dimensional
space-time and has great advantage over other methods of evaluating
anomaly such as calculating directly the Feynman diagram and observing the
non-invariance of path integral measure proposed by Kazuo Fujikawa
\cite{fuji}. Actually, this approach provided a basis for Michael
Green and John Schwarz to find the celebrated anomaly cancelation
mechanism in
 superstring theory \cite{green} and hence to some extent
 promoted the first superstring revolution.\\

 The application of the descent equations for constructing the
 Donaldson polynomial in terms of topological Yang-Mills theory
 works like `` killing two birds with one stone ". The chain of
 descent equations, $\delta Q^{p}_{2n-p-1}=-dQ_{2n-p-2}^{p+1}$,
 shows that $\delta Q^{p}_{2n-p-1}$
 ($0\leq p\leq 2n-1$), the BRST transformation of a certain $(2n-p-1)$-degree
 differential form with ghost number $p$ on a
 $2n$-dimensional manifold $M^{2n}$, equals  to $dQ_{2n-p-2}^{p+1}$, the exterior
 differential of the one-degree-less differential form
 $Q_{2n-p-2}^{p+1}$ with ghost number $p$. Therefore, in topological
 Yang-Mills theory, if we can  find  a set of $Q_{2n-p}^{p-1}$'s composed
 of the field variables and choose  their integrals over
 $(2n-p)$-homological cycles $\Sigma_{2n-p}$ on the space-time manifold
 as physical observables, then these BRST
invariant physical observables not only describe a direct
correspondence between the BRST cohomological class in the Hilbert
space of the field theory and the homological class on the
space-time manifold, but also reflect that the BRST cohomology can
reduce to a cohomology over the Yang-Mills instanton moduli space
once the non-zero modes of field variables have been integrated out.
Witten chose the second Chern class $\mbox{Tr}(F\wedge F)$ as  the
top differential form and constructed other three differential forms
with the application of the BRST transformations \cite{witt1}. These
four differential forms composed of the field functions  constitute
a chain of descent equations. The integrals of these four
differential forms over homological cycles $\Sigma_p$ ($0\leq p\leq
3$) on the space-time manifold are BRST invariants and depend only
on the homological class $\Sigma_p$, and hence constitute a set of
fundamental physical observables to produce topological invariants
on the space-time manifold. According to the topological selection
rule on the physical process induced by the instanton tunneling
effect \cite{hooft1}, one can use the product of  above four
fundamental observables to construct a set of physical observables
whose ghost numbers equal to the dimension of instanton moduli space
$d({\cal M})$ \cite{witt1}. Then the expectation values of these
physical observables, after integrating out the non-zero modes,
indeed yield the Donaldson polynomial invariants constructed through
the mapping from the homology group on the space-time manifold to
the homological
group over the compactified Yang-Mills instanton moduli space.\\

Witten's topological Yang-Mills theory  has not only provided  a
physical approach to reproduce the Donaldson invariant, but also
initiated a new type of quantum field theory \cite{witt1}.  Usual
quantum field theories have an infinite number of degrees of
freedoms and its Hilbert space is infinite dimensional.  But local
quantum states in the Hilbert space of a topological field theory
are only vacuum states, and it is exactly solvable to some extent.
In physics topological quantum field theory describes a special
phase of field theory in which there exist only topological (or
no-local) quantum excitations. It might be helpful for understanding
quantum gravity since a graviton can arise as the Goldstone boson
corresponding to the spontaneous breaking of general covariance in
topological quantum field theory, despite this symmetry breaking is
hard to realize. From the end of 1980s to the early stage of 1990s,
the study on topological quantum field theory is a hot topic for
both theoretical physicists and mathematicians. However, it should
emphasize that the topological Yang-Mills theory approach is not
much helpful for the practical calculation on the Donaldson
invariant. This can be easily understood from physical
consideration. First, $N=2$ supersymmetric Yang-Mills theory is
strongly coupled at low-energy and there is no a feasible way to
solve the theory; Second, the integration over the instanton moduli
space suffers from infrared divergence due to the large size of
instanton (or large volume of instanton moduli space), and this
brings obstacles on the explicit calculation on the Donaldson
invariant. This stagnant situation lasted until the early middle of
1990s.

\section{Electric-magnetic Duality, Seiberg-Witten Monopole
Equation and Donaldson
Invariant}

 In 1994 Nathan Seiberg and Edward Witten made a  breakthrough in understanding
 $N=2$ non-perturbative supersymmetric Yang-Mills
 theory in four dimensions \cite{seiwit}. They
 synthesized a number of  nonperturbative  methods of  solving
 supersymmetric gauge theory and physical phenomena occurred in
 the supersymmetric gauge theory including supersymmetric instanton
 calculus, spontaneous breaking of gauge symmetry, the holomorphy
 imposed by supersymmetric Ward
 identities and  quantum anomaly of superconformal symmetry.
 Especially, they made use of electric-magnetic duality
 conjecture, which  is a notion that  had left out in the cold for
 a long time. With the accumulative  application of these methods,
Seiberg and Witten successfully  determined the structure of
vacuum moduli space for $N=2$ supersymmetric Yang-Mills theory in
the Coulomb phase and the corresponding quantum Wilsonian effective action
including all of the perturbative and non-perturbative quantum
corrections. They further gave a quantitative explanation to the
mechanisms for confinement and chiral symmetry breaking.  This is
the first time in the history of quantum field theory that
 an exact solution for a strongly coupled quantum field
theory in four dimensions has been obtained. The electric-magnetic
duality conjecture plays a crucial role in attaining the solution.
The electric-magnetic dual transformation in the vacuum moduli
space of the Coulomb phase of $N=2$ supersymmetric Yang-Mills
theory  is represented as a subgroup of the discrete group
$SL(2,Z)$, which is generated by transformation matrices  (called
monodromy) around the singularities in the vacuum moduli space. In
the following we give a brief introduction to the history of
electric-magnetic duality in
order to understand Seiberg and Witten's work.\\

Electric-magnetic duality conjecture has a long history and it came
out not long after the birth of  modern physics.  As early as the
beginning of 1930s,  Dirac realized the asymmetry of electric and
magnetic field sources in the Maxwell equations and hence proposed
the notion of magnetic monopole \cite{dirac}, which is a particle
carrying magnetic charge, just like a charged particle carrying
electric charge. The Maxwell equations with magnetic charge sources
present an elegant electric-magnetic duality: when the electric and
magnetic charges change their roles with each other, the electric
and magnetic fields also exchange (up to a sign). Dirac further
found that when one considers quantum mechanics of an electrically
charged particle moving in the magnetic field produced by a
monopole, in order to achieve single-valuedness of the wave
function, the product of the electric charge of the particle with
the magnetic charge of the monopole must be a multiple of $2\pi$,
this is the celebrated Dirac quantization condition. In
electrodynamics the electric charge is directly related to
electromagnetic coupling, so the Dirac quantization condition
implies that electric-magnetic duality  can exchange a strongly
coupled electric theory with a weakly coupled magnetic theory and
\emph{vice versa}. However, a quantum field theory with magnetic
monopole had  not been  found until 40 years later. In 1974, Gerard
't Hooft \cite{hooft2} and A.M. Polyakov \cite{poly1} independently
found a $SU(2)$ gauge field theory that mediates scalar field
interaction and contains a self-interacting scalar potential, which
breaks the $SU(2)$ gauge symmetry break to $U(1)$ symmetry
spontaneously, can have a topological soliton solution with finite
energy. This solution presents the feature of a magnetic monopole
and now is called the 't Hooft-Polyakov monopole. At large distance,
the 't Hooft-Polyakov monopole behaves exactly as the Dirac
monopole. By the way, the gauge field theory with monopole solution
was actually the bosonic part of a field model proposed by Howard
Georgi and Sheldon Glashow \cite{glashow}, which describes the
suppression of the physical process involving leptonic neutral currents,
and hence is usually called the Georgi-Glashow model. When the
energy of the theory takes a  minimal finite value and the
self-interaction of scalar field is very weak, the explicit analytic
form of the 't Hooft-Polyakov monopole solution can be worked out.
This is the so-called Bogomol'nyi-Prasad-Sommerfield (BPS) magnetic
monopole \cite{bogo}. The mass of the BPS monopole is proportional
to its magnetic charge. On the other hand, since the
self-interacting scalar potential causes the spontaneous breaking of
$SU(2)$ symmetry to $U(1)$, two electrically charged components of
gauge field acquire mass through the Higgs mechanism and the masses
proportional to electric charges they carry. Naturally, both the
masses of BPS monopole and gauge field measured in terms of the
vacuum expectation value of the scalar field since it is the only
parameter with mass dimension. Several years after the discovery of
't Hooft-Polyakov monopole, Claus Montonen and David Olive observed
the symmetry between the mass spectrums of gauge particle and BPS
monopole, which are  the fundamental quantum excitation and
topological soliton of the George-Glashow model, respectively. Hence
they boldly proposed that the Georgi-Glashow model should have a
physically equivalent magnetic dual description, in which the BPS
monopole should play the role of a gauge field and the massive
vector boson in the broken phase of the Georgi-Glashow model becomes
an ``electric" monopole \cite{mool}. In particular, when the Dirac
quantization condition is taken into account, the gauge couplings of
these two dual theories should be opposite.  Later, Witten
\cite{witt8} considered the strong charge-conjugate and parity
symmetry violation effect in the George-Glashow model and introduced
a topological  $\theta$-term proportional to the instanton number.
This term causes the integer-valued magnetic charge to have a shift
proportional to the $\theta$ parameter consistent with the Dirac
quantization  condition \cite{witt8}. Consequently, the duality
operation extends from $Z_2$ to $SL(2,Z)$ transformation.\\

However, the Montonen-Olive duality conjecture for the
Georgi-Glashow model has several fatal problems. First,  the vector
bosons and magnetic monopole as dual particles have different spins;
Second, quantum correction  can usually modify the classical
potential \cite{cole}, and hence corrects the mass of gauge
particles acquired from  the spontaneous breaking of gauge symmetry
happened at classical level. Third, the gauge coupling in a quantum
field theory changes along with the energy scale. Finally, since
electric-magnetic duality involves two theories with opposite
couplings, it is almost not possible to verify this duality since
usually there is no way to solve a strongly coupled theory. Two
years after Montonen and Olive proposed the conjecture, Hugo Osborn
from the Cambridge University found that these drawbacks can be
refrained in the $N=4$ supersymmetric Yang-Mills theory \cite{osbo}.
Thus electric-magnetic duality should exist in  a gauge theory with maximal
global supersymmetry. After Osborn's work, the notion of
electric-magnetic duality had been laid aside and almost
neglected in field theory.\\

This situation lasted until early 1990s,  the Indian string theorist
Ashoke Sen found a duality existing in the heterotic string theory
compactified on a six-dimensional torus \cite{asen}. The resultant
theory after compactification is a four-dimensional string theory
with $N=4$ supersymmetry, and its low-energy effective theory is
$N=4$ supergravity coupled with $N=4$ supersymmetric Yang-Mills
theory. Therefore, this duality is the Montonen-Olive-Osborn duality
in heterotic string theory. In the latter half of 1994, Seiberg and
Witten found that $N=2$ supersymmtric Yang-Mills theory in the
Coulomb phase at low-energy also presents an electric-magnetic
duality \cite{seiwit}, with which an exact Wilsonian effective action
can be determined, and led to a great stir among theoretical
physicists. This work had also rung the bell for the second
superstring revolution. However, it should be emphasized that the
duality found by Seiberg and Witten  is different from the
Montonen-Olive-Osborn duality in $N=4$ supersymmetric Yang-Mills
theory. The latter is somehow a self-duality, the magnetic monopole
lies in the  $N=4$ vector supermultiplet, and the classical action
of magnetic dual theory has the same form as the original
``electric" theory. But for the Seiberg-Witten duality, the magnetic
monopole lies in $N=2$ matter supermultiplet (hypermultiplet), that
is, the monopole belongs to a matter superfield. This
electric-magnetic duality is actually close to the original duality
conjecture proposed by Dirac. To summarize, according to Seiberg and
Witten's work, the low-energy Wilsonian effective action of $N=2$
supersymmetric Yang-Mills theory in the Coulomb phase,
composed of perturbative quantum correction and  non-perturbative
contribution from instanton tunneling effect, is dual to a $N=2$
supersymmeytric $U(1)$ gauge theory weakly coupled with a $N=2$
magnetic monopole hypermultiplet.
These two dual theories describe the same physics.\\

Shortly after the discovery of the Seiberg-Witten duality in $N=2$
supersymmetric Yang-Mills theory, Witten  applied  it to the
calculation on the Donaldson invariants \cite{witt5}. As introduced
before, the Donaldson invariant is a physical observable of the
twisted $N=2$ supersymmetric Yang-Mills theory, which can be
expressed as an integration over the Yang-Mills instanton moduli
space. Viewed from physics side, they are physical amplitudes of
fermionic zero modes in the instanton background. According to
electric-magnetic duality, these physical observables can be
equivalently calculated in the dual magnetic theory. The calculation
on the physical observables becomes much more tractable since the
dual theory is a $N=2$  Abelian supersymmetric gauge theory.
Especially,  the physical obervable in the  twisted dual magnetic
theory can only come from the integration over the moduli space of
magnetic monopole solution. The weak-coupling of the magnetic dual
theory means that it is sufficient to achieve the Donaldson
invariant by considering only the moduli space of classical magnetic
monopole, which is now called the Seiberg-Witten monopole equation.
Therefore, the calculation on the Donaldson invariant in the
magnetic dual theory converts into counting the number of
independent solutions to the Seiberg-Witten monopole equation.
Overall, the study on the differential topology of a simply
connected four-manifold in terms of a relativistic quantum field
theory had finally achieved a successful outcome.

\section{Other Applications of Quantum Field Theory in Differential Geometry}

There are more examples about the application of quantum field
theory in the study on topology, differential geometry and
algebraic geometry. Some of them are list as the following.

\begin{itemize}

\item The Morse inequalities in the Morse theory, which is an
alternative and powerful approach to investigate topological
structure of a manifold by studying the critical points of a
function defined on the manifold, was derived through the tunneling
effect in $0+1$-dimensional supersymmetric nonlinear sigma model
\cite{witt4}.

\item The Atiyah-Singer index theorem, which states the analytical
index of an elliptic differential operator (i.e., the number of
solutions to a differential equation) should  equal to the
topological index of the manifold on which the operator is defined,
was verified by calculating the Witten index \cite{witt9} in
$0+1$-dimensional supersymmetric nonlinear sigma model
\cite{lag,fried}.

\item The elliptic cohomology was proposed as a generalization of $K$-theory
in algebraic geometry and correspondingly, the elliptic genus should
be related to the elliptic cohomology  in the same way as the Dirac
index is related to $K$-theory.  Witten suggested that the role in
elliptic cohomology analogous to the Dirac operator in $K$-theory
should be played by the supercharge of an $N=(1,1)$ supersymmetric
non-linear sigma model in $1+1$ dimensions \cite{wittad},  provided
that the left-moving fermion is assigned to the Neveu-Schwarz
boundary condition and the right-moving fermion to a Ramond-Ramond
boundary condition. The elliptic genus is precisely the index of the
supercharge operator, just like the $\widehat{A}$-genus being the
index of the Dirac operator.

\item The Jones polynomials which classify knots and links
topologically in three-dimensional space are identical to quantum
Wilson loops of Chern-Simons topological gauge
theory in three dimensions\cite{witt2}.

\item  The Thurston conjecture which can give a  topological
classification on three-dimensional manifolds could be studied by a
certain three-dimensional gravity theory \cite{kuns}.

\end{itemize}

Since the middle of 1980s there has arisen an upsurge in studying
superstring theory since in physics it can give a consistent
theoretical description on quantum gravity and possibly provide a
framework for unifying four fundamental interactions. The elementary
particles, which are represented by quantum field in gauge field
theory, now appear as various vibrational modes on open string and
closed string, which are termed according to  their geometrical
shapes looking like a segment or a circle of thread, respectively.
The dynamics of string theory is described by certain
two-dimensional $N=1$ supersymmetric nonlinear sigma models but with
conformal symmetry. The rising of superstring theory has brought
about many new mathematical tools to apply in physics. For example,
the dynamical symmetry of string theory on world-sheet is conformal
symmetry in two dimensions, which is described by infinite
dimensional Virasovo and Kac-Moody algebras \cite{2cft}; The
calculation on multi-loop amplitude of string theory needs the
knowledge about the Riemann-Roch theorem on the Riemannian surface
and Teichm\"{u}ller space etc. \cite{hoph}. The conformal symmetry
of quantum string theory requires the background space-time must be
ten-dimensional. To get to the anticipated physics in four-dimensional
space-time, we need to perform a compactification from
ten to four dimensions, and the compactified six-dimensional space
must be a Calabi-Yau manifold \cite{styau,stcom}. In the middle of
1990s the discovery on the string soliton -- D-brane led to a
breakthrough in understanding non-perturbative string theory. The
formerly discovered five string theories can be connected together
by dualities and unified into a so-called $M$-theory living in an
eleven-dimensional space-time. Even now people has not understood
much about $M$-theory such as the fundamental principle behind it
and its fundamental degrees of freedom. The only confirmed fact is
that its low-energy limit should be the $N=1$ supergravity in eleven
dimensions. $D$-brane plays a crucial role in establishing string
dualities and it  carries topological charges with respect to the
antisymmetric fields produced by quantum excitations of the
world-sheet fermion in closed string theory under the periodic
boundary condition. Witten pointed out that the topological charge
carried by $D$-brane  can be classified by the $K$-theory in fibre
bundle theory \cite{witten7}. String theory has occupied so many
advanced mathematics, and reversely, it can also be considered as a
tool to study differential geometry and algebraic geometry. For
instance, the mirror symmetry between two Calabi-Yau manifolds is a
pure geometrical property and string theory  greatly facilitates the
study on mirror symmetry. At the beginning of 1990s, Witten
constructed two types of topological strings (called $A$- and
$B$-models) by `` twisting " two-dimensional $N=2$ supersymmetric
non-linear sigma model in two different ways, and the search on a
pair of mirror manifolds converts into calculating and comparing
physical observables of two topological strings \cite{witt3}. In
recent years, theoretical physicists and mathematicians are
interested in topological string theory in the background of
$D$-instanton, this may helpful for exploring the mirror symmetry
between a pair of the generalized Calabi-Yau manifolds
\cite{hitchin}. In a colloquium on the prospects of mathematics in
the 21st century,  the leading mathematician in geometric analysis and
mathematical physics and the Field medalist, Professor Shing-Tung
Yau at the Harvard University,
 pointed out that `` the developments in string theory have
successfully unified some important sectors of differential
geometry, algebraic geometry, group representation theory, number
theory and topology " and he predicted that `` the grand unification
in mathematics could be bred  from and born out of the grand unified
field theory in physics " \cite{styau2}. These facts indicate that
string theory has much greater impacts on the development in
mathematics.

\section{Summary}

We have introduced the physical idea how a relativistic quantum
field theory has been used to derive the Donaldson invariants in
differential geometry. The aim is to use this fact as an example to
illustrate how quantum gauge theory, which is a theoretical
framework describing the interaction among elementary particles, can
be applied to study differential topology. It shows that once a
physical notion has a mathematical counterpart, a physical framework
can be used as a tool to study mathematics. Both quantum field
theory and string theory have deep roots in differential geometry,
algebraic geometry and group theory,  and most of notions in these
two theoretical frameworks are just another appearances (or physical
representations) of certain mathematical objects. Thus it is no
wonder that they are becoming an alternative way to develop new
mathematics. Therefore, the interrelation between mathematics and
physics are stepping into a new era.
Mathematics and physics are two
oldest disciplines in the scientific civilization of mankind, both
born out of the cognition on natural world. They have finally melted
into with each other after more than two hundred years' relatively
independent developments, despite that there were some intersections
in the past times. Nowadays more and more mathematicians and
theoretical physicists are learning from each other to propose
mathematical conjectures and prove new
theorems.\\


\vspace{5mm}

\noindent \textbf{Acknowledgments:} This work is partially supported
by the Natural Sciences and Engineering Research Council of Canada.
It is based on a colloquium on \emph{Quantum Field Theory and Modern
Differential Geometry} given at the Department of Mathematics and
Statistics of the University of Guelph; several seminars on
\emph{Topological Quantum Field Theory and Topological String}
delivered in the String Theory Group at the Department of Physics of
National Taiwan University and especially a series of lectures  on
\emph{Quantum Field Theory and Differential Geometry} given at the
Department of Mathematics of National Taiwan University. I am
obliged to Professor I-Hsun Tsai at the Department of Mathematics in the
National Taiwan University for numerous discussions on the relevant
knowledge in Differential Geometry and Algebraic Geometry. I am
indebted to Professors Gabor Kunstatter and Randy Kobes at the
Department of  Physics in the University of Winnipeg for financial
support. Finally, I would like to thank Professor Masud Chaichian at
the Department of Physics in the University of Helsinki and
Professor Roman Jackiw at the Department of Physics in the
Massachusetts Institute of Technology for their continuous
encouragements and comments.


\begin{thebibliography}{99}

\bibitem{yangm} C.N. Yang and R. Mills,
\emph{Conservation of Isotopic Spin and Isotopic Gauge Invariance},
Phys. Rev. \textbf{96} (1954) 191.

\bibitem{lubk} E. Lubkin, \emph{Geometric Definition of Gauge Inariance},
Ann. Phys. \textbf{23} (1963) 233.

\bibitem{wuyang} T.T. Wu and C.N. Yang,
\emph{Concept of Nonintegrable Phase Factors and Global Formulation
of Gauge Fields}, Phys. Rev. \textbf{D12} (1975) 3845; \emph{Dirac
Monopole Without Strings: Monopole Harmonics}, Nucl.Phys.B107 (1976)
365.

\bibitem{dirac} P.A.M. Dirac, \emph{Quantized Singularities in the
Eletromagnetic Field},  Proc. Roy. Soc., London, \textbf{A133}
(1931) 60.

\bibitem{bpst} A.A. Belavin, A. M. Polyakov, A.S. Schwarz and Yu.S.
Tyupkin, \emph{Pseudoparticle Solutions of the Yang-Mills
Equations}, Phys. Lett. \textbf{B59} (1975) 85.

\bibitem{chern1} S.S. Chern, \emph{On the Curvature Integral in A
Riemannian Manifold}, Ann. Math. \textbf{46} (1945) 674.

\bibitem{witten6} E. Witten,
\emph{Some Exact Multi-Instanton Solutions of Classical Yang-Mills
Theory}, Phys. Rev. Lett. \textbf{38} (1977) 121.

\bibitem{cofa1} E. Corrigan and D.B. Fairlie,
\emph{Scalar Field Theory and Exact Solutions to a Classical SU(2)
Gauge Theory}, Phys. Lett.\textbf{ B67} (1977) 69; E. Corrigan, D.B.
Fairlie, R.G. Yates and P. Goddard, \emph{The Construction of
Selfdual Solutions to SU(2) Gauge Theory}, Commun. Math. Phys.
\textbf{58} (1978) 223.

\bibitem{jackiw4} R. Jackiw, C. Nohl and C. Rebbi \emph{Conformal Properties of
Pseudoparticle Configurations}, Phys. Rev. \textbf{D15} (1977) 1642.

\bibitem{jackiw2} R. Jackiw and C. Rebbi,
\emph{Conformal Properties of a Yang-Mills Pseudoparticle}, Phys.
Rev. \textbf{D14} (1976) 517.

\bibitem{schw} A.S. Schwarz, \emph{On Regular Solutions of Euclidean
Yang-Mills Equations}, Phys. Lett. \textbf{B67} (1977) 172.

\bibitem{atiya1} M.F. Atiyah, N.J. Hitchin and I.M. Singer,
Deformations of Instantans, Proc. Nat. Acad. Sci. \textbf{74} (1977)
2662

\bibitem{jackiw3} R. Jackiw and C. Rebbi,
\emph{Degrees of Freedom in Pseudoparticle Systems}, Phys. Lett.
\textbf{B67} (1977) 189.

\bibitem{bclee} L.S. Brown, R.D. Carlitz and C. Lee,
\emph{Massless Excitations in Pseudoparticle Fields}, Phys. Rev.
\textbf{D16} (1977) 417.

\bibitem{guth} C.W. Bernard, N.H. Christ, A.H. Guth and E.J.
Weinberg, \emph{Instanton Parameters for Arbitrary Gauge Groups},
Phys. Rev. \textbf{D16} (1977) 2967.

\bibitem{atiya2} M.F. Atiyah, N.J. Hitchin and I.M. Singer,
\emph{Self-duality in Four-dimensional Riemannian Geometry}, Proc.
Roy. Soc. Lond. \textbf{A362} (1978) 425.


\bibitem{ward} R.S. Ward, \emph{On Selfdual Gauge Fields},
Phys. Lett. \textbf{A61} (1977) 81; M.F. Atiyah and R.S. Ward,
\emph{Instantons and Algebraic Geometry}, Commun. Math. Phys.
\textbf{55} (1977) 117; M.F. Atiyah, N.J. Hitchin, V.G. Drinfeld and
Yu.I. Manin, \emph{Construction of Instantons}, Phys. Lett.
\textbf{A65} (1978) 185.

\bibitem{penrose} R. Penrose and M.A.H. MacCallum,
\emph{Twistor Theory: An Approach to the Quantization of Fields and
Space-time}, Phys. Rept. \textbf{6} (1972) 241.

\bibitem{simons}J.P. Bourguignon, H.B. Lawson, and J. Simons,
 \emph{Stability and Gap Phenomena for Yang-Mills Fields},
 Proc. Nat. Acad. Sci. \textbf{76} (1979) 1550.

\bibitem{jackiw1} R. Jackiw and C. Rebbi,
 \emph{Vacuum Periodicity in a Yang-Mills Quantum Theory},
 Phys. Rev. Lett. \textbf{37} (1976) 172.

\bibitem{gross1} C.G. Callan, Jr., R.F. Dashen and D.J. Gross,
\emph{The Structure of the Gauge Theory Vacuum}, Phys. Lett.
\textbf{B63} (1976) 334.

\bibitem{hooft1} G. 't Hooft, \emph{Computation of the Quantum Effects
due to a Four-Dimensional Pseudoparticle}, Phys. Rev. \textbf{D14}
(1976) 3432.

\bibitem{gross2} C.G. Callan, Jr., R.F. Dashen and D.J. Gross,
\emph{Toward a Theory of the Strong Interactions}, Phys.Rev.
\textbf{D17} (1978) 2717.

\bibitem{anomaly} S.L. Adler, \emph{Axial Vector Vertex in Spinor
Electrodynamics}, Phys. Rev. \textbf{177} (1969) 2426; J.S. Bell,
and R. Jackiw, \emph{A PCAC Puzzle: $\pi_0\rightarrow 2\gamma$ in
the $\sigma$-Model}, \textbf{A60} (1969) 47.

\bibitem{indano} R. Jackiw and C. Rebbi, \emph{Spinor Analysis of Yang-Mills
Theory}, Phys. Rev. \textbf{D16} (1977) 1052; N.K. Nielsen and B.
Schroer, \emph{Axial Anomaly and Atiyah-Singer Theorem}, Nucl. Phys.
\textbf{B127} (1977) 493; C. Callias, \emph{Index Theorems on Open
Spaces}, Commun. Math. Phys. \textbf{62} (1978) 213; M. Ninomiya and
C.-I Tan, \emph{Axial Anomaly And Index Theorem For Manifolds With
Boundary}, Nucl. Phys. \textbf{B257} (1986) 199.

\bibitem{ins1} I. Affleck, M. Dine and  N. Seiberg,
\emph{Supersymmetry Breaking by Instantons}, Phys. Rev. Lett.
\textbf{51} (1983) 1026; \emph{Dynamical Supersymmetry Breaking in
Supersymmetric QCD}, Nucl. Phys. \textbf{B241} (1984) 493.

\bibitem{ins2} V.A. Novikov, M.A. Shifman, A.I. Vainshtein
and V.I. Zakharov, \emph{Instantons In Supersymmetric Theories},
Nucl. Phys. \textbf{B223} (1983) 445; \emph{Exact Gell-Mann-Low
Function of Supersymmetric Yang-Mills Theories from Instanton
Calculus}, Nucl. Phys. \textbf{B229} (1983) 381; \emph{Supersymmetry
Transformations of Instantons}, Nucl. Phys. \textbf{B229} (1983)
394; \emph{Instanton Effects in Supersymmetric Theories}, Nucl.
Phys. \textbf{B229} (1983) 407; \emph{Supersymmetric Instanton
Calculus (Gauge Theories with Matter)}, Nucl. Phys. \textbf{B260}
(1985) 157.

\bibitem{ins3} D. Amati, G.C. Rossi, G. Veneziano,
\emph{Instanton Effects in Supersymmetric Gauge Theories}, Nucl.
Phys. \textbf{B249} (1985) 1; \emph{Massive SQCD And The Consistency
Of Instanton Calculations}, Nucl. Phys. \textbf{B263} (1986) 591;
\emph{Nonperturbative Aspects in Supersymmetric Gauge Theories},
Phys. Rept. 162 (1988) 169.

\bibitem{ins4} N. Dorey, V.V. Khoze and  M.P. Mattis,
\emph{Multi-instanton Calculus in N=2 Supersymmetric Gauge Theory},
Phys. Rev. \textbf{D54} (1996) 2921; \emph{Multi-instanton Calculus
in N=2 Supersymmetric Gauge Theory: 2. Coupling to Matter}, Phys.
Rev. \emph{D54} (1996) 7832; \emph{Supersymmetry and the
Multi-instanton Measure}, Nucl. Phys. \textbf{B513} (1998) 681;  N.
Dorey, T.J. Hollowood, V.V. Khoze and M.P. Mattis,
\emph{Supersymmetry and the Multi-Instanton Measure II: From N=4 to
N=0}, Nucl. Phys. \textbf{B519} (1998) 470.

\bibitem{witt1} E. Witten, \emph{Topological Quantum Field Theory},
Commun. Math. Phys. \textbf{117} (1988) 353.

\bibitem{seiwit} N. Seiberg and E. Witten,
\emph{Monopoles, duality and chiral symmetry breaking in N=2
supersymmetric QCD}, Nucl. Phys. \textbf{B431} (1994) 484;
\emph{Electric-magnetic duality, monopole condensation, and
confinement in N=2 supersymmetric Yang-Mills theory}, Nucl. Phys.
\textbf{B426} (1994) 19.

\bibitem{wess} J. Wess and B. Zumino,
\emph{A Lagrangian Model Invariant under Supergauge
Transformations}, Phys. Lett. \textbf{B49} (1974) 52; \emph{
Supergauge Invariant Extension of Quantum Electrodynamics}, Nucl.
Phys. \textbf{B78} (1974) 1.

\bibitem{colman2} S. Coleman and J. Mandula, \emph{All Possible
Symmetries of the S Matrix},
Phys. Rev. \textbf{159} (1967) 1251.


\bibitem{salam} A. Salam and J.A. Strathdee,
\emph{On Superfields and Fermi-Bose Symmetry}, Phys. Rev.
\textbf{D11} (1975) 1521; S. Ferrara, J. Wess and B. Zumino, Phys.
Lett. B51 (1974) 239.

\bibitem{aff} I. Affleck, \emph{On Constrained Instantons},
Nucl. Phys. \textbf{B191} (1981) 429.

\bibitem{witt5} E. Witten, \emph{Monopoles and Four-Manifolds},
Math. Res. Lett. \textbf{1} (1994) 769; S\emph{upersymmetric
Yang-Mills Theory on a Four-Manifold}; J. Math. Phys. \textbf{35}
(1994) 5101; \emph{On S duality in Abelian Gauge Theory}, Selecta
Math. \textbf{1} (1995) 383.

\bibitem{dona} S.K. Donaldson,
\emph{An Application of Gauge Theory to Four-dimensional Topology},
J. Diff. Geom. \textbf{18} (1983) 279; \emph{Instantons And
Geometric Invariant Theory}, Commun. Math. Phys. \textbf{93} (1984)
453; \emph{Polynomial Invariants for Smooth Manifolds}, Topology
\textbf{29} (1990) 257.

\bibitem{free} M. Freedman, \emph{The Topology of Four-diemnsional
Manifolds}, J. Diff. Geom. \textbf{17} (1982) 357.

\bibitem{ghost} L.D. Faddeev and V.N. Popov,
\emph{Feynman Diagrams for the Yang-Mills Field}, Phys. Lett.
\textbf{B25} (1967) 29.

\bibitem{brs} C. Becchi, A. Rouet an R. Stora,
\emph{Renormalization of Gauge Theories}, Ann. Phys. \textbf{98}
(1976) 287.

\bibitem{tyu} I.V. Tyupin, \emph{Gauge Invariance in Field Theory
and in Statistical Physics in the Operator Formalism}, Lebedev
Preprint FIAN No. 39 (1975), unpublished.

 \bibitem{atijones} M.F. Atiyah and J.D.S. Jones,
 \emph{Topological Aspects of Yang-Mills Theory},
 Commun. Math. Phys. \textbf{61} (1978) 97.

\bibitem{atisi} M.F. Atiyah, I.M. Singer, \emph{Dirac Operators
Coupled to Vector Potentials}, Proc. Nat. Acad. Sci. \textbf{81}
(1984) 2597.

\bibitem{wuzee2} Y.-S. Wu and  A. Zee,
\emph{Abelian Gauge Structure inside Nonabelian Gauge Theories},
Nucl. Phys. \textbf{B258} (1985) 157.

\bibitem{reina1} G. Falqui and C. Reina,
\emph{BRS Cohomology and Topological Anomalies}, Commun. Math. Phys.
\textbf{102} (1985) 503.

\bibitem{houzhang} B.-Y. Hou and Y.-Z. Zhang,
\emph{Cohomology in Connection Space, Family Index Theorem
 and Abelian Gauge Structure}, J. Math. Phys. \textbf{28} (1987) 1709;
Y.-Z. Zhang, \emph{Covariant Anomaly and Cohomology in Connection
Space}, Phys. Lett. \textbf{B219} (1989) 439.

\bibitem{mieg} J. Thierry-Mieg, \emph{Geometrical Reinterpretation of
Faddeev-Popov Ghost Particles and BRS Transformations}, J. Math.
Phys. 21 (1980) 2834. .

 \bibitem{bono} L. Bonora and  P. Cotta-Ramusino, \emph{Some Remarks on
BRS Transformations, Anomalies and the Cohomology of the Lie Algebra
of the Group of Gauge Transformations}, Commun. Math. Phys.
\textbf{87} (1983) 589.

\bibitem{eguchi} See for example, T. Eguchi, P.B. Gilkey and A.J. Hanson,
\emph{Gravitation, Gauge Theories and Differential Geometry}, Phys.
Rept. \textbf{66} (1980) 213, Subsect. 5.5.

\bibitem{stora2} R. Stora, \emph{Algebraic Structure and Topological Origin of
Anomalies}, Seminar given at Cargese Summer Institute:
\emph{Progress in Gauge Field Theory}, Cargese, France, September,
1983.



\bibitem{zumi2} B. Zumino, \emph{Chiral Anomalies and Differential Geometry},
Lectures given at Les Houches Summer School on Theoretical Physics,
Les Houches, France, August, 1983, published in \emph{Relativity,
Groups and Toplogy II} edited by  B.S. DeWitt and R. Stora
(North-Holland, Amsterdam, 1984); \emph{Cohomology of Gauge Groups:
Cocycles and Schwinger Terms}, Nucl. Phys. \textbf{B253} (1985) 477.


\bibitem{zumi} J. Ma\~{n}es, R. Stora and B. Zumino, \emph{Algebraic Study
of Chiral Anomalies}, Commun. Math. Phys. \textbf{102} (1985) 157;
W. Bardeen and B. Zumino, \emph{Consistent and Covariant Anomalies
in Gauge and Gravitational Theories}, Nucl. Phys. \textbf{B244}
(1984) 421.


\bibitem{japan} T. Kugo and I. Ojima, \emph{Local
Covariant Operator Formalism of Nonabelian Gauge Theories and Quark
Confinement Problem}, Prog. Theor. Phys. Suppl. \textbf{66}
 (1979) 1.

\bibitem{wezu2} J. Wess and B. Zumino, \emph{Consequences of anomalous
Ward identities}, Phys. Lett. \textbf{B37} (1971) 95.

\bibitem{zuwze} B. Zumino, Y.-S. Wu and A. Zee, \emph{Chiral Anomalies,
Higher Dimensions and Differential Geometry}, Nucl. Phys.
\textbf{B239} (1984) 477.

\bibitem{bard} W.A. Bardeen, \emph{Anomalous Ward Identities in
Spinor Field Theories}, Phys. Rev. 184 (199) 1848.


\bibitem{dvmv} M. Dubois-Violette, M. Talon and C.M. Viallet,
\emph{New Results on BRS Cohomology in Gauge Theory}, Phys. Lett.
\textbf{B158} (1985) 231;  \emph{BRS Algebras: Analysis of the
Consistency Equations in Gauge Theory}, Commun. Math. Phys.
\textbf{102} (1985) 105.

\bibitem{manes} J. Ma\~{n}es and B. Zumino, \emph{Nontriviality Of
Gauge Anomalies},  Nuffield Workshop (1985).

\bibitem{fadd2} L.D. Faddeev, \emph{Operator Anomaly for the Gauss Law},
Phys. Lett. \textbf{B145} (1984) 81.

\bibitem{jackwco} R. Jackiw, \emph{3-Cocycle in Mathematics and Physics},
Phys. Rev. Lett. \textbf{54}  (1985) 159.

\bibitem{wuzee3} Y.-S. Wu and A. Zee, \emph{Cocycles And Magnetic
Monopoles}, Phys. Lett. \textbf{B152} (1985) 98; B. Grossman,
\emph{The Meaning of the Third Cocycle in the Group Cohomology of
Nonabelian Gauge Theories}, Phys. Lett. \textbf{B160} (1985) 94;
S.G. Jo, \emph{Commutators in an Anomalous Nonabelian Chiral Gauge
Theory}, Phys. Lett. \textbf{B163} (1985) 353; \emph{Commutator Of
Gauge Group Generators In Nonabelian Chiral Theory},
 Nucl. Phys. \textbf{B259} (1985) 616; Y.-Z. Zhang,
 \emph{Realization of Three Cocycle of Gauge Group in Hamiltonian Dynamics}
  Phys. Lett. \textbf{B189} (1987) 149;
D. Levy, \emph{The Failure Of The Jacobi Identity for Free Fermionic
Currents
 and Its Relation to the Axial Anomaly}, Nucl.Phys. \textbf{B282} (1987) 367.


\bibitem{alga} L. Alvarez-Gaume and E. Witten,  \emph{Gravitational Anomalies},
Nucl. Phys. \textbf{B234} (1964) 269.


\bibitem{fuji} K. Fujikawa,  \emph{Path Integral Measure for Gauge Invariant Fermion
Theories}, Phys. Rev. Lett. \textbf{42} (1979) 1195; \emph{Path
Integral for Gauge Theories with Fermions}, Phys. Rev. \textbf{D21}
(1980) 2848.

\bibitem{green} M.B. Green and J.H. Schwarz,  \emph{Anomaly Cancellation in
Supersymmetric D=10 Gauge Theory and Superstring Theory}, Phys.
Lett. \textbf{B149} (1984) 117; \emph{The Hexagon Gauge Anomaly in
Type I Superstring Theory},
 Nucl. Phys. \textbf{B255} (1985) 93.


\bibitem{hooft2} G. 't Hooft, \emph{Magnetic Monopoles in Unified Gauge
Theories}, Nucl. Phys. \textbf{B79} (1974) 276.

\bibitem{poly1} A.M. Polyakov, \emph{Particle Spectrum in the Quantum Field Theory},
JETP Lett. \textbf{20} (1974) 194.

\bibitem{glashow} H. Georgi and S. L. Glashow,
 \emph{Unified Weak and Electromagnetic Interactions without Neutral
 Currents}, Phys. Rev. Lett. \textbf{28} (1972) 1494.

 \bibitem{bogo} M.K. Prasad and C.M.
Sommerfield,  \emph{An Exact Classical Solution for the 't Hooft
Monopole and the Julia-Zee Dyon},  Phys. Rev. Lett. \textbf{35}
(1975) 760; E.B. Bogolmol'nyi,  \emph{Stability of Classical
Solutions}, Sov. J. Nucl. Phys. \textbf{24} (1976) 449.

\bibitem{mool} C. Montonen and  D.I. Olive,\emph{ Magnetic Monopoles
as Gauge Particles?}, Phys. Lett. \textbf{B72} (1977) 117.

\bibitem{witt8} E. Witten, \emph{Dyons of Charge $e \theta/(2 \pi)$},
Phys. Lett. \textbf{B86} (1979) 283.


\bibitem{cole} S.R. Coleman and E. Weinberg, \emph{Radiative Corrections
as the Origin of Spontaneous Symmetry Breaking}, Phys. Rev.
\textbf{D7} (1973) 1888.


\bibitem{osbo} H. Osborn, \emph{Topological Charges for N=4 Supersymmetric
Gauge Theories and Monopoles of Spin 1}, Phys. Lett. \textbf{B83}
(1979) 321.

\bibitem{asen} A. Sen, \emph{Electric-Magnetic Duality in String
Theory}, Nucl. Phys. \textbf{B404} (1993) 109; Q\emph{uantization of
Dyon Charge and Electric-Magnetic Duality in String Theory}; Phys.
Lett. \textbf{B303} (1993) 22; \emph{$SL(2,Z)$ Duality and
Magnetically Charged Strings}; Int. J. Mod. Phys. \textbf{A8} (1993)
5079; \emph{Magnetic Monopoles, Bogomolny Bound and $SL(2,Z)$
Invariance in String Theory}, Mod. Phys. Lett. \textbf{A8} (1993)
2023.


\bibitem{witt4} E. Witten, \emph{Supersymmetry and Morse Theory},
J. Diff. Geom. \textbf{17} (1982) 661.



\bibitem{witt9} E. Witten, \emph{Dynamical Breaking of Supersymmetry},
Nucl. Phys.\textbf{ B188} (1981) 513; \emph{Constraints on
Supersymmetry Breaking}, Nucl. Phys. \textbf{B202} (1982) 253.

\bibitem{lag} L. Alvarez-Gaume,  \emph{Supersymmetry and the Atiyah-Singer Index
Theorem}, Commun. Math. Phys. \textbf{90} (1983) 161; \emph{A Note
On The Atiyah-Singer Index Theorem}, J. Phys. \textbf{A16} (1983)
4177.

\bibitem{fried} D. Friedan and P. Windey, \emph{Supersymmetric Derivation of t
he Atiyah-Singer Index and the Chiral Anomaly}, Nucl. Phys.
\textbf{B235} (1984) 395.


 \bibitem{wittad} E. Witten, \emph{Elliptic Genera and Quantum Field Theory},
 Commun. Math. Phys. \textbf{109} (1987) 525.


\bibitem{witt2} E. Witten, \emph{Quantum Field Theory and the Jones
Polynomial}, Commun. Math. Phys. \textbf{121} (1989) 351.


\bibitem{kuns} J. Gegenberg and G. Kunstatter  \emph{Using 3-D stringy gravity
to understand the Thurston conjecture}, Class. Quant. Grav.
\textbf{21} (2004) 1197.


\bibitem{2cft} A.A. Belavin, A.M. Polyakov and A.B. Zamolodchikov,
\emph{Infinite Conformal Symmetry in Two-Dimensional Quantum Field
Theory}, Nucl. Phys. \textbf{B241} (1984) 333.

\bibitem{hoph} E. D'Hoker and D.H. Phong, \emph{The Geometry of String
Pertubation Theory}, Rev. Mod. Phys. \textbf{60} (1988) 917.

\bibitem{styau} S.-T. Yau, \emph{Calabi's Conjecture and some new results in
algebraic geometry}, Proc. Nat. Acad. Sci. \textbf{74} (1977) 1798.

\bibitem{stcom} P. Candelas, G.T. Horowitz, A. Strominger,
and E. Witten, \emph{Vacuum Configurations for Superstrings}, Nucl.
Phys. \textbf{B258} (1985) 46;  Strominger, and E. Witten, \emph{New
Manifolds for Superstring Compactification}, Commun. Math. Phys.
\textbf{101} (1985) 341.

\bibitem{witten7} E. Witten, \emph{D-branes and K theory},
JHEP \textbf{9812} (1998) 019; \emph{Overview of K-theory Applied to
Strings}, Int. J. Mod. Phys. \textbf{A16} (2001) 693;  G.W. Moore
and E. Witten, \emph{Selfduality, Ramond-Ramond Fields and
K-theory}, JHEP \textbf{0005} (2000) 032;

\bibitem{witt3} E. Witten, \emph{Topological Sigma Models},
Commun. Math. Phys. \textbf{118} (1988) 411; \emph{Mirror Manifolds
and Topological Field Theory}, in \emph{Mirror Symmetry} edited by
S.T. Yau, page 121, {\tt hep-th/9112056}; \emph{Chern-Simons Gauge
Theory as a String Theory}, Prog. Math. \textbf{133} (1995) 637.

\bibitem{hitchin} N.J. Hitchin, \emph{Generalized Calabi-Yau
Manifolds}, Quart. J. Math. Oxford  Ser. \textbf{54}  (2003) 281,
{\tt math/0209099}


\bibitem{styau2} S.T. Yau, \emph{The Prospects for Mathematics in the
Twenty-first Century}, colloquium given in the  National Center of
Theoretical Sciences at the National Tsinghua University, Hsinchu,
Tainwan (in Chinese).







\end{thebibliography}
\end{document}